\newcommand{\version}{June 29, 2016}
\documentclass[a4paper,oneside,11pt,pdftex]{article}
\pdfoutput=1
\usepackage[bindingoffset=0.0cm,textheight=22.6cm,hdivide={*,15.9cm,*}, vdivide={*,22cm,*}]{geometry}
\usepackage{amsmath,amsfonts,amssymb}
\usepackage{mathtools}
\usepackage[pdftex]{graphicx}
\usepackage[T1]{fontenc}
\usepackage[utf8]{inputenc}
\usepackage{lmodern}
\usepackage{tensind}
\tensordelimiter{?}
\IfFileExists{dsfont.sty}
	{\usepackage{dsfont}
         \let\mathbb=\mathds
         \newcommand{\id}{\mathds{1}}}
	{\typeout{Package dsfont.sty was not found, using alternative macros.}
         \let\mathds=\mathbb
         \newcommand{\id}{\mbox{1 \kern-.59em \textrm{l}}}}

\let\Id=\id
\usepackage[pdftex,hyperref,svgnames]{xcolor}
\usepackage[pdftex,bookmarksnumbered=true,breaklinks=true,colorlinks=true,linktocpage=true,linkcolor=MediumBlue,citecolor=ForestGreen,urlcolor=DarkRed]{hyperref}
\makeatletter
\AtBeginDocument{
  \hypersetup{
    pdftitle = {\@title},
    pdfauthor = {Daniel N. Blaschke, Fran\c{c}ois Gieres,
M\'eril Reboud, and Manfred Schweda}
  }
}
\makeatother

\usepackage[numbers,square,comma,sort&compress]{natbib}

\usepackage{etoolbox}
\newtoggle{grouprefs}
\toggletrue{grouprefs}
\iftoggle{grouprefs}{

\newcommand{\dymarskynakayama}{dymarskynakayama}
\newcommand{\becchi}{becchi}

\newcommand{\montesinos}{montesinos}
\newcommand{\einstein}{einstein}

\newcommand{\hehlhammond}{hehlhammond}

\newcommand{\stora}{stora-1980ies}
\newcommand{\kastlerstora}{kastler-stora}
\newcommand{\GirardiFerrara}{GirardiFerrara}
\newcommand{\lazzarini}{lazzarini}
\newcommand{\deser}{deser}
\newcommand{\straumann}{straumann}
}
{
\newcommand{\dymarskynakayama}{Nakayama:2013is,Dymarsky:2013pqa}
\newcommand{\becchi}{Becchi:1974xu,Becchi:1974md,Becchi:1975nq}
\newcommand{\montesinos}{Montesinos:2006th,GamboaSaravi:2003aq}
\newcommand{\einstein}{Einstein:1915ca,Einstein:1916a}
\newcommand{\hehlhammond}{Hehl:1976kj,Hammond:2002rm}
\newcommand{\stora}{Stora:1987qy,Stora:1995er}
\newcommand{\kastlerstora}{Kastler:1986mj,Stora:2005dq,Stora:1993zw,Stora:1996ip}
\newcommand{\GirardiFerrara}{Girardi:1985hf,Ferrara:1993yj}
\newcommand{\lazzarini}{Lazzarini:1988rv,Lazzarini:1989gp}
\newcommand{\deser}{Deser:1969wk,Deser:2009fq}
\newcommand{\straumann}{Straumann-book,Straumann:2000ke}
}

\makeatletter

 \@addtoreset{equation}{section}
 \makeatother

\newcommand{\EMT}{EMT }

\newcommand{\vp}{\varphi}
\newcommand{\mg}{\textbf{\texttt{g}}}
\newcommand{\mq}{q}
\newcommand{\hg}{\hat U}
\newcommand{\gro}{U}

\newcommand{\varep}{\varepsilon}

\newcommand{\lie}{\mathtt{\mathbf{g}}}
\newcommand{\br}{\mathbb{R}}

\newcommand{\txt}[1]{\textrm{#1}}

\newcommand{\Div}{\textrm{div}}

\newcommand{\Tr}{\textrm{Tr}}

\newcommand{\secref}[1]{section~\ref{#1}}		
\newcommand{\appref}[1]{appendix~\ref{#1}}		
\newcommand{\co}[2]{\left[#1,#2\right]}					
\newcommand{\pa}{\partial}						
\newcommand{\diff}[2]{\frac{\pa #1}{\pa #2}}				
\newcommand{\ri}{\textrm{i}}						
\newcommand{\re}{\textrm{e}}						
\newcommand{\m}{\mu}
\newcommand{\n}{\nu}

\newcommand{\s}{\sigma}

 \newcommand{\cH}{\mathcal{H}} 
  \newcommand{\cL}{\mathcal{L}}

\newcommand{\nn}{\nonumber}

\newcommand{\preU}{\prescript{\gro}{}\!} 

\newcommand{\Boxed}[1]{\setlength\fboxsep{0.4em}\boxed{\ #1 \ }}

\title{\texorpdfstring{\begin{flushright}
\vspace*{-2cm}{\small LYCEN-2016-04 \\[-1ex] LA-UR-16-23055}
       \end{flushright}\vspace{2em}}{}%
The Energy-Momentum Tensor(s)
 in Classical Gauge Theories}
\date{\version}

\author{Daniel N. Blaschke\footnotemark[1]~, Fran\c{c}ois Gieres\footnotemark[2]~,
M\'eril Reboud\footnotemark[2] \footnotemark[3]~,
 \\and Manfred Schweda\footnotemark[4]}

\begin{document}
\renewcommand{\thepage}{\roman{page}}


\maketitle
\thispagestyle{empty}
\begin{center}
\renewcommand{\thefootnote}{\fnsymbol{footnote}}
\vspace{-0.3cm}
\footnotemark[1]Los Alamos National Laboratory,\\Los Alamos, NM, 87545, (USA)\\[0.3cm]
\footnotemark[2]
Institut de Physique Nucl\'eaire de Lyon, \\
Universit\'e de Lyon, Universit\'e Claude Bernard Lyon 1 and CNRS/IN2P3,\\Bat. P. Dirac, 4 rue Enrico Fermi,
F-69622-Villeurbanne (France)
\\[0.3cm]
\footnotemark[3] Ecole Normale Sup\'erieure de Lyon, \\
46 all\'ee d'Italie,
F-69364-Lyon CEDEX 07 (France)\\[0.3cm]
\footnotemark[4]Institute for Theoretical Physics, Vienna University of Technology\\Wiedner Hauptstra\ss e 8-10, A-1040 Vienna (Austria)
\\[0.5cm]
\ttfamily{E-mail: dblaschke@lanl.gov, gieres@ipnl.in2p3.fr, meril.reboud@ens-lyon.fr, mschweda@tph.tuwien.ac.at}
\end{center}

\vspace{1.0em}
\begin{center}
\emph{Dedicated to the memory of Raymond Stora}
\\[0.5em]
\emph{who made profound and lasting contributions to field theory and to symmetries in physics}
\end{center}

\vspace{1.0em}
\begin{abstract}
We give an introduction to, and review of, the energy-momentum tensors in classical gauge field
theories in Minkowski space, and to some extent also in curved space-time.
For the canonical energy-momentum tensor of non-Abelian gauge fields and of matter fields coupled
to such fields, we present a new and simple improvement procedure
based on gauge invariance for constructing a gauge invariant, symmetric energy-momentum  tensor.
The relationship with the Einstein-Hilbert tensor following from the coupling to a gravitational field is also
discussed.
\end{abstract}

\newpage
 \renewcommand{\thepage}{\arabic{page}}
\setcounter{page}{1}
\tableofcontents


\section{Introduction}

The role and impact of symmetries in modern physics can hardly be overstated.
As H.~Weyl~\cite{Weyl:1952} put it: \emph{``As far as I see,
all a priori statements in physics have their origin in symmetry''.}
Some salient features are the following ones~\cite{Gieres:1997iw}:
\begin{itemize}
\itemsep=3pt
\item
The laws of nature are possible realizations of the symmetries of nature.
\item
The basic quantities or \emph{building blocks} of physical
theories are often defined and classified
by virtue of symmetry considerations,
 e.g. elementary particles and relativistic fields.
\item
The \emph{general structure} of physical theories is largely
determined by the underlying invariances. In particular,
the \emph{form of interactions} is strongly res\-tricted
by geometric symmetries (relativistic covariance)
and the gauge symmetries essentially fix all fundamental
interactions (electro-weak, strong and gravitational forces).
\end{itemize}
A pillar of classical mechanics and field theory is given by the  two theorems that E.~Noether
established in 1918~\cite{Noether:1918zz}.
They describe in quite general terms the consequences of the invariance
of an action functional under a Lie group of global or
local symmetry transformations, respectively, extending and generalizing some
special cases which had previously been investigated, in particular by F.~Klein in relationship with general relativity,
e.g. see~\cite{Olver, Kosmann, Sundermeyer:2014kha, Zuber:2013rha} and references therein.
These two theorems can be applied to systems with a finite number of degrees of freedom (mechanics) as well as to systems
with an infinite number of degrees of freedom (field theory),
both in their relativistic or non-relativistic versions, e.g. see reference~\cite{Banados:2016zim}
for numerous examples. They can also be generalized to superspace~\cite{Ogievetsky:1978js, Lang:1978ws} or to non-commutative space (see~\cite{Zahn:2003bt} and references therein).

\paragraph{Plan of the paper:}
In the present note we focus, for  classical YM (Yang-Mills) theories in $n$-dimensional  Minkowski space, on the EMT
(\emph{energy-momentum tensor}, also referred to as \emph{stress-energy tensor}
or \emph{stress tensor} for short):
the components $T^{\m \n}$ of the EMT can be interpreted as follows, e.g.
see reference~\cite{Ohanian:2013}.
$T^{00}$ represents the \emph{energy density,} $T^{0i}$ the  \emph{$i$-momentum density (or energy flux density)}
and $T^{ij}$ the  \emph{$i$-momentum flux density in the $j$-direction.}
Regarding the field theoretical system as a
collection of particles, we can also interpret $T^{ii} \equiv p$ as the \emph{pressure} and $(T^{ij})$
with $i\neq j$  as the \emph{shear stress.}
In particular, we will study here the so-called canonical EMT whose components $T_{\txt{can}} ^{\m \n}$
 represent the conserved current densities
which are associated (by virtue of Noether's first theorem) to the  invariance of the action
under space-time translations. The corresponding conserved charges
$P^{\nu} \equiv \int_{\br^{n-1}} d^{n-1} x \, T_{\txt{can}}^{0 \nu}$ define the total energy and momentum of the physical system.

As is well known (for instance for Maxwell's theory), the tensor  $T_{\txt{can}} ^{\m \n}$ is
neither symmetric nor gauge invariant in general and thereby needs to be ``improved''.
This is traditionally realized by the ``symmetrization procedure of Belinfante''~\cite{Belinfante:1939, Belinfante:1940}
which relies on the spin angular momentum density,
but this method does not work straightforwardly in the case where matter fields are minimally coupled to  a gauge field~\cite{Grensing:2013}.
After a short introduction to the subject and problematics in subsection~\ref{sec:IntrodEMT},
we will show in subsection~\ref{sec:diffimpEMT} that the improvement can be realized
in a simple manner for pure gauge fields or for
interacting gauge and matter fields by taking into account the local gauge
invariance\footnote{In the older literature  (e.g. see reference~\cite{Ryder:1996}),
the \emph{local,} i.e. space-time dependent, gauge transformations are
often referred to as \emph{gauge transformations of the second kind}
as opposed to \emph{gauge transformations of the first kind}, i.e. \emph{global}
gauge transformations (labeled by constant parameters).
To avoid a topological connotation, some authors also use the terminology \emph{flexible} versus
\emph{rigid} symmetry transformations for transformations involving $x$-dependent symmetry parameters versus constant ones.
By \emph{gauge invariance} we shall always mean local gauge invariance.}.
A conceptually quite different approach consists of coupling the gauge and matter fields to gravity and
deducing the so-called \emph{Einstein{-}Hilbert EMT} in Minkowski space from the metric EMT on curved space.
This approach is  outlined for YM theories in section~\ref{sec:gravity}
and it is readily shown that the different results in Minkowski space coincide with each other.
Concerning the latter point we
should mention that more general and abstract approaches have been considered
in the literature (we refer in particular to the systematic study~\cite{Forger:2003ut} based on the earlier work~\cite{GotayMarsden}),
but we hope that the elementary discussion of the different aspects presented here is useful both as a short introduction to,
and as an overview of, the subject.
While our paper is devoted to the classical theory, we conclude with some remarks
on how symmetries impact the quantum theory, i.e. on Ward identities, and more generally on the related subjects to which R.~Stora
made substantial contributions.

\paragraph{General framework of classical Lagrangian field theory in Minkowski space:}

Our  conventions and general assumptions are as follows.
 We choose natural units so that
$c \equiv 1 \equiv \hbar$.
The points of $n$-dimensional Minkowski space are labeled by $x \equiv (x^\mu)_{\mu = 0, 1, \dots, n-1}
\equiv (t, \vec x \, )
\in \br^n$ and the signature
of the Minkowski metric $(\eta_{\mu \nu})$ is chosen to be $(+, -, \dots, -)$.
We will deal only with YM theories, i.e. multiplets of complex scalar or Dirac fields which are minimally coupled
to the YM field $A^\m (x) \equiv A^\m _a (x ) T_a$ (see
appendix~\ref{sec:GFTnutshell}
for the notation and for a summary of gauge theories).
Thus, generically we have a collection
$\varphi \equiv (\vp_r)_{r = 1, \dots, N}$
of \emph{classical relativistic fields in Minkowski space}
which are assumed to
vanish (together with their derivatives) sufficiently fast at spatial infinity.
The dynamics of fields is specified by
a \emph{local, Poincar\'e invariant action functional} $S[\varphi ] \equiv \int d^nx \, {\cal L} (\varphi , \pa_{\mu} \varphi)$
which involves a Lagrangian density ${\cal L}$ having the following properties.
We consider closed systems which means~\cite{LandauLifschitz2} that the Lagrangian ${\cal L}$ does not explicitly depend  on the space-time coordinates, i.e. the $x$-dependence of $ {\cal L}$ is only due to the $x$-dependence of
$\varphi$ and $\pa_{\mu} \varphi$.
Moreover, we assume that ${\cal L}$ depends at most on first-order derivatives:
the associated equations of motion
\begin{align}
0 = \frac{\delta S}{\delta \varphi} =
 \frac{\pa {\cal L}}{\pa \vp}
- \pa_{\mu} \left( \frac{\pa {\cal L}}{\pa (\pa_{\mu} \vp)} \right)
\,, \label{eq:LAGeom}
\end{align}
are then at most of second order.
We concentrate on the Lagrangian formulation of field theory
and refer to the work~\cite{Hanson:1976cn, IP}
for a discussion of the EMT of gauge theories
within the Hamiltonian framework.
Moreover in section~\ref{sec:gravity}, we discuss the generalization to curved space-time manifolds.

\section{EMT's for gauge theories in Minkowski space}
\label{sec:EMTlag}

After recalling the definition of  the canonical
EMT~\cite{LandauLifschitz2, Grensing:2013, Duncan:2012},
we consider the example of pure YM theories
(while referring whenever necessary to the summary of non-Abelian gauge theories
given in appendix~\ref{sec:GFTnutshell}).
Thereafter, we discuss the  general properties of the EMT
(local conservation law, symmetry, gauge invariance, tracelessness)
as well as the addition of superpotential terms
to the canonical EMT: The aim of these additions is to ``improve'' the characteristics of the canonical EMT,
 so that it becomes symmetric and gauge invariant (and traceless for scale invariant Lagrangians) if it does not have these properties.

\subsection{Canonical \EMT and its properties}\label{sec:IntrodEMT}

\subsubsection{The canonical EMT and the associated conserved charges}

\paragraph{Noether's first theorem:}
By virtue of Noether's first theorem, the invariance of the action $S[\varphi ]$
under translations $x^\n \to x^{\prime \nu} = x^\n + a^\n$ implies the existence of
current densities $(j_{\txt{can}}^\mu)^\n$ satisfying the conservation law
$\pa_\m (j_{\txt{can}}^\mu)^\n =0$ for
$\n \in \{ 0, 1, \dots, n-1 \}$.
The conserved second-rank tensor $T_{\txt{can}}^{\mu \nu} \equiv (j_{\txt{can}}^\mu)^\n$
is referred to as the \emph{canonical EMT} field
and with the notation
$\delta x^\m = a^\m , \,  \delta \vp = - a^\m \pa_\m \vp$  (where $| a^\m | \ll 1$)
we have
\begin{align}
\Boxed{
0 = \frac{\delta S}{\delta \vp} \, \delta \vp + \pa_\m j^\m
}
\qquad \txt{with} \quad
\Boxed{
j^\m  \equiv  - T^{\m \n} _{\txt{can}} \; a_\n
}
\qquad
\Boxed{
T_{\txt{can}} ^{\m \n} \equiv  \frac{\pa {\cal L}}{\pa (\pa_{\mu} \vp)} \; \pa^\n \vp  -\eta^{\m \n} {\cal L}
}
\, .
\label{eq:canEMT}
\end{align}
In this expression for the tensor field $T_{\txt{can}} ^{\m \n}$ and in similar expressions, the sum over all fields is
implicitly understood (e.g. the sum over $\phi$ and $\phi^*$ in the case of a complex scalar field $\phi$).
The \emph{local conservation law} $\pa_{\mu} T_{\txt{can}}^{\mu \nu} =0$
holds by virtue of the equations of motion of $\vp$ and its explicit verification
amounts to a one-line derivation of $ T_{\txt{can}}^{\mu \nu} $ \cite{LandauLifschitz2}:
\[
\pa_\m (\eta^{\m \n} {\cal L} ) = \pa^\n {\cal L} \, =
\underbrace{\, \frac{\pa {\cal L}}{\pa \vp} \,}_{=  \, \pa_\m \big(  \frac{\pa {\cal L}}{\pa (\pa_{\mu} \vp)} \big) }
\; \pa^\n \vp
+
 \frac{\pa {\cal L}}{\pa (\pa_{\mu} \vp)} \;
 \underbrace{\pa^\n (\pa_\m \vp  )}_{= \, \pa_\m (\pa^\n \vp ) } = \pa_\m \Big(  \frac{\pa {\cal L}}{\pa (\pa_{\mu} \vp)} \; \pa^\n \vp
 \Big)
 \, .
\]

If we assume as usual that the fields vanish sufficiently fast at spatial infinity, then the boundary term
$\int d^{n-1} x \,\pa_{i} T_{\txt{can}}^{i \nu}$ vanishes and we have \emph{conserved ``charges''}
\begin{align}
\Boxed{
P^{\nu} \equiv \int_{\br^{n-1}} d^{n-1} x \, T_{\txt{can}}^{0 \nu}
}
\label{eq:IntConsLaw}
\end{align}
associated with the conserved densities \eqref{eq:canEMT}.
The vector $(P^\n)$  represents the \emph{total energy-momentum  of the fields}
and generates space-time translations of the fields $\varphi$ in  classical and in quantum theory.
(Some subtleties appear for gauge theories
due to the presence of so-called constraint equations
following from the gauge invariance of the Lagrangian, see references~\cite{Hanson:1976cn, IP}.)
In particular $P^0$ coincides with the
\emph{canonical Hamiltonian function} $H$ since
$T^{00}_{\txt{can}}=\diff{\cL}{\dot\vp} \, \dot\varphi-\cL=\cH$.
The fact that $\int_V d^{n-1} x \, T_{\txt{can}}^{0 \nu}$
represents the components of a Lorentz vector for any space region $V \subset \br^{n-1}$
(a result which is also known as \emph{von Laue's theorem}~\cite{Griffiths:2012})
is not quite obvious, but can be verified~\cite{Schroder:1990ri} by using
the assumptions that $\pa _\m T_{\txt{can}} ^{\m \n} =0$ and that the functions $T_{\txt{can}} ^{\m \n} $
vanish sufficiently fast at the boundary of $V$.
We note that the integral~\eqref{eq:IntConsLaw} is performed over a hypersurface of $\br ^n$ given by $x^0$ constant
which can be replaced by a generic $(n-1)$-dimensional
space-like hypersurface $\Sigma$
with fields vanishing on its boundary $\pa \Sigma$.

\paragraph{Other derivations:}
We note that the canonical Noether currents associated to
geometric symmetries can also be obtained~\cite{Itzykson:2005}
by considering \emph{local} symmetry transformations
without coupling the fields to a gravitational field
(and similarly for internal symmetries
without a coupling of matter fields to a gauge field).
This approach may be referred to as \emph{Gell-Mann-L\'evy procedure}~\cite{Kosyakov:2007qc}
since it goes back to the classic work of Gell-Mann and L\'evy on
the $\sigma$-model~\cite{GellMann:1960np}.

\subsubsection{Superpotential terms}

Apart from an overall numerical factor, there are other ambiguities in the definition of a canonical current.
In fact, for the  canonical EMT we can always add a so-called  \emph{superpotential term,}
i.e. the divergence of an antisymmetric tensor, so as to pass over to a
 so-called \emph{improved EMT} $T_{\txt{imp}}^{\mu \nu} $ defined by\footnote{One may regret that some  monographs (e.g.~\cite{Ryder:1996}) refer to~\eqref{eq:imprEMT}
as ``the canonical EMT'' since $\chi^{\mu \rho\nu} $ and thereby $T_{\txt{imp}}^{\mu \nu} $
are neither unique nor naturally given.}
\begin{align}
 \Boxed{
T_{\txt{imp}}^{\mu \nu}  \equiv T_{\txt{can}}^{\mu \nu}  + \pa_{\rho} \chi^{\mu \rho\nu}
}
\, , \qquad \txt{with} \ \;
 \Boxed{
\chi^{\mu \rho\nu} = - \chi^{\rho \mu \nu}
}
\label{eq:imprEMT}
\,.
\end{align}
The antisymmetry of $\chi^{\mu \rho\nu}$ in its first two indices ensures that
$\pa_{\rho} \chi^{\mu \rho\nu}$ is identically conserved, hence
$T_{\txt{imp}}^{\mu \nu} $ \emph{is conserved} as well.
The derivative in the superpotential term entails that
\emph{one has the same conserved charge as for}
 $T_{\txt{can}}^{\mu \nu}$
provided $\chi^{0 i\n}$ decreases for $r \equiv |\vec x \, | \to \infty$ faster than $1/r^{n-2}$
in $n$ space-time dimensions.
The freedom~\eqref{eq:imprEMT} may be used to give a different form
(eventually with a greater physical significance)
to the EMT, various examples
for this ``improvement procedure'' (i.e. judicious choices of superpotentials) being given below.

\subsubsection{The canonical EMT for pure YM theory and its general properties}

\paragraph{Canonical expression:}
By way of example, for pure YM theory in $n$ dimensions, the translation invariance of the YM action~\eqref{eq:invact},
i.e. $ {\cal L}_g (F) \equiv  - \frac{1}{4c_2} \, \Tr \, (F^{\mu \nu} F_{\mu \nu} )$,
yields
\begin{align}
T_{\txt{can}}^{\mu \nu} =
 \frac1{c_2} \, \Tr \, ( -  F^{\mu \rho}  \, \pa^{\nu}  A_\rho   + \frac{1}{4} \, \eta^{\mu \nu} F^{\rho \sigma}  F_{\rho \sigma} )
\, .
\label{eq:canEMTYM}
\end{align}
According to Noether's first theorem, this EMT is locally conserved, i.e. $\pa_{\mu} T_{\txt{can}}^{\mu \nu} =0$,
by virtue of the field equation
$0= D_\n F^{\n \m} \equiv \pa_\n F^{\n \m}+\ri q\co{A_\n}{F^{\n \m}} $ where $q$ denotes the  non-Abelian charge
(cf. \appref{sec:GFTnutshell}). As for any EMT
some other properties are also of interest: one may wonder
whether it is \emph{gauge invariant} (since we are dealing with gauge theories),
whether it is \emph{symmetric} in its indices, whether it is \emph{traceless}
and whether it gives rise to a \emph{positive energy density}.
As a matter of fact, the EMT~\eqref{eq:canEMTYM} enjoys none of these properties.
The physical and mathematical ideas behind these four properties are the following ones,
ignoring for the moment the coupling to gravity which we address in the next subsection.

\paragraph{On the gauge invariance of the EMT:}
The lack of gauge invariance of the components of the
classical EMT  is unacceptable  since these
physical quantities  are measurable;
and so is the energy-momentum $\int_V d^{n-1}x \, T^{0\nu}_{\txt{can}}$
(contained in a domain $V\subset \br^{n-1}$ of finite volume) which is not on-shell gauge invariant
(i.e. not gauge invariant for fields $\vp$ satisfying the classical equations of motion).

\paragraph{On the symmetry of the EMT:}
If the \EMT $T_{\txt{can}}^{\mu \nu}$ is not symmetric
on-shell, then the canonical angular momentum tensor $M_{\txt{can}}^{\mu \nu \rho}$ and  the EMT $T_{\txt{can}}^{\mu \nu}$ are not related
in the same way as angular  momentum and momentum are related in classical mechanics,
i.e. by a relation of the form ``the components of $M$ are the moments of $T$'',
\begin{align}
M^{\mu \rho \sigma} = x^{\rho} T^{\mu \sigma} - x^{\sigma} T^{\mu \rho}
\label{eq:AMT}
\,.
\end{align}
Indeed for a tensor $M^{\mu \rho \sigma} $ of this form,
the conservation law $\pa_{\mu} M^{\mu \rho \sigma} = 0$ only holds if the tensor
 $T^{\mu \nu}$ is symmetric on-shell:
\begin{align}
0 = \pa_{\mu} M^{\mu \rho \sigma} =\pa_{\mu}( x^{\rho} T^{\mu \sigma} - x^{\sigma} T^{\mu \rho})
= T^{\rho \sigma} - T^{\sigma \rho}
\label{eq:ConsAMT}
\,,
\end{align}
where we used the fact that $\pa_{\mu} T^{\mu \nu} =0$ holds for solutions of the equations of motion.
Thus, \emph{if the canonical EMT is not symmetric, then the canonical angular momentum tensor
does not have the `mechanical' form~\eqref{eq:AMT}, rather we have an extra spin angular
momentum term $s^{\m \rho \sigma}$,}
 \begin{align}
M_{\txt{can}}^{\mu \rho \sigma} = x^{\rho} T_{\txt{can}}^{\mu \sigma} - x^{\sigma} T_{\txt{can}}^{\mu \rho}
+ s^{\mu \rho \s}
\, , \quad \txt{with} \ \; s^{\mu \rho \sigma} \equiv \frac{\pa {\cal L}}{\pa (\pa_{\mu} \vp)}
\, (\frac{1}{2} \, \Sigma^{\rho \sigma} ) \, \vp = - s^{\mu \sigma \rho}
\,.
\label{eq:canAMT}
\end{align}
Here, $\frac{\ri}{2} \, \Sigma^{\rho \sigma} \equiv  d(M_{\rho \sigma})$
is a $N$-dimensional representation $d$ of a basis $(M_{\rho \sigma})$ of the  Lie algebra associated to the Lorentz group, e.g.
 $ d(M_{\rho \sigma})  =0$ for a scalar field, $ d(M_{\rho \sigma}) \equiv  M_{\rho \sigma}$ for a vector field $(A^\m)$ and
 $  d(M_{\rho \sigma})  \equiv   \frac{1}{8} \, [ \gamma_{\rho} , \gamma_{\sigma} ]$
 for a Dirac field
 ($\gamma_0 , \dots, \gamma_{d-1}$ denoting matrices satisfying the Clifford algebra relation
 $\{ \gamma_\rho , \gamma_\sigma \} = 2 \eta_{\rho \sigma} \Id$).

\paragraph{On the tracelessness of the EMT:}
Consider a field theoretic model which is invariant
under rescalings $x \leadsto x' = \re ^{\rho} x$ (where $\rho$ is a constant real parameter),
e.g. pure YM theory in four dimensions
or the theory of a real massless scalar field $\phi$.
For such a model,
one expects that the canonical EMT can be improved
so as to obtain a ``new improved'' EMT $T^{\m \n} _{\txt{conf}}$
which is conserved, symmetric and
traceless~\cite{Callan:1970ze, Jackiw:2011vz}:
the dilatation current
$(j^\m )$ associated to scale invariance
is then given by the moments of the EMT  and its local conservation law
(reflecting the scale invariance) is tantamount to the tracelessness of the EMT:
\begin{align}
j^\m = x_\n T^{\m \n}_{\txt{conf}}
\, ,
\qquad \txt{hence} \ \quad
\pa_\m j^\m = T^{\m}_{\txt{conf}\,\m} =0
\, .
\label{eq:CLTS}
\end{align}
This tracelessness of the EMT plays an important role in conformal field theories~\cite{DiFrancesco:1997nk}.
Here, we only note that for a single, real, massless scalar field $\phi$
in $n$-dimensional Minkowski space,  the
\emph{new improved} or \emph{Callan-Coleman-Jackiw EMT}~\cite{Callan:1970ze} reads
\begin{align}
T^{\m \n}_{\txt{conf}} \equiv
 T^{\m \n}_{\txt{can}}
 - \xi(n) \, (\pa^\m \pa^\n - \eta^{\m \n} \Box ) \phi^2
 \qquad \txt{with} \ \; \Box \equiv \pa^\m \pa_\m
 \quad \txt{and} \ \; \xi (n) \equiv \frac{1}{4} \,  \frac{n-2}{n-1}
 \label{eq:confEMT}
\,.
\end{align}

\paragraph{On the positivity of the energy density:}

An interesting question is whether or not the total energy $P^0$ is positive
or, more importantly (the energy being only determined up to an additive constant in the absence of gravity),
whether \emph{the total energy is bounded from below.}
In fact, if this is not the case, then one can extract an infinite amount of energy
from the physical system.
 If the energy density $T^{00}$
 satisfies the positivity condition $T^{00} \geq 0$ in all inertial frames, then the energy $P^0$ is positive
 in all inertial frames
 if $|\vec P | \leq P^0$, i.e. equivalently
 if the vector $P\equiv (P^0, \vec P \, )$ is time-like and future-directed\footnote{Under a Lorentz boost with velocity
 $\vec v$, the component $P^0$ goes over to $P^{\prime 0} = \gamma (P^0 - \vec P \cdot \vec v \, )$ with
 $\gamma \equiv (1 - \vec v^{\, 2})^{-1/2}$, hence $P^0 \geq 0$ does not imply $P^{\prime 0} \geq 0$
 without the given assumption.}~\cite{Thirring:1997}.
 The latter condition is satisfied if the EMT satisfies the so-called \emph{dominant energy condition}~\cite{Sexl:2001}
 stipulating that the energy current ${\cal E}^\m \equiv {T^\m}_\n  u^\n$ is time-like
 and future-directed for every observer with velocity $(u^\m )$ (e.g. consider $(u^\m ) = (1 , \vec 0 \, )$).
 The energy conditions reflect the principles of relativity and play an important role in general relativity
 where relations of this type are an essential ingredient for establishing general results like the no hair theorem,
 the laws of black hole thermodynamics or the singularity theorems
 of Penrose and Hawking
 which predict the occurrence of singularities in the
 universe~\cite{Hawking:1973uf,Senovilla:2014gza, Wald:1984, Poisson-book}.

\subsubsection{Improvements of the canonical EMT and definition of the physical EMT}\label{sec:assessment}

\paragraph{Improvements of the canonical EMT:}
In view of the unpleasant and problematic properties (lack of symmetry and of gauge invariance) of the canonical \EMT
some improvements of the canonical expression \eqref{eq:canEMT} have been looked for.
The first one is due to Belinfante~\cite{Belinfante:1939}
and represents an astute procedure for constructing a \emph{symmetric} \EMT
out of the canonical expression --- see section~\ref{sec:BelProc}.
In~\secref{sec:impgaugetheory}  we will show
that, for non-Abelian gauge theories, the same results can be obtained by taking into account
the \emph{gauge invariance} of the Lagrangian.
(Equivalently, one can already exploit this gauge invariance in establishing Noether's first theorem
for the translational invariance so as to obtain right away a gauge invariant (and symmetric) EMT,
see references~\cite{Eriksen:1979vq, Takahashi:1985dt, Munoz:1996wp}
as well as~\cite{\montesinos}; we
also note that a symmetric EMT can be obtained in establishing Noether's theorem if one
considers both translations and Lorentz transformations in an appropriate way~\cite{Barut:1981}.)
We mentioned already that for scale invariant theories a further improvement, which is due to Callan, Coleman and Jackiw~\cite{Callan:1970ze},
consists of constructing  a \emph{traceless} \EMT  out of Belinfante's symmetric tensor.

\paragraph{Definition of the physical EMT:}
The canonical EMT and its improvements
represent different localizations for the energy and momentum, and one may wonder whether
there is a preferred or a correct localization.
An answer to this
question can be provided by coupling the matter fields (including gauge fields)
to the gravitational field.
In fact, in general relativity the \EMT
represents the local source for the gravitational field in Einstein's field equations very much like the
electric current density
represents the source for the electromagnetic field in Maxwell's equations.
Thus, in curved space-time, one defines the so-called \emph{metric EMT} for the  gauge or matter field
as the functional derivative of the
gauge/matter field action $S[\vp, g_{\m \n}]$ with respect to the metric tensor field $(g_{\m \n})$,
see section~\ref{sec:gravity} below.
Accordingly, this curved space EMT is generally covariant, covariantly conserved, symmetric and gauge invariant.
The so-called \emph{Einstein-Hilbert EMT} $T^{\m \n} _{\txt{EH}} $ in Minkowski space which is obtained
from the metric EMT by setting
$g_{\m \n} = \eta _{\m \n} $ is then conserved, symmetric and gauge invariant by construction.

As was pointed out by Rosenfeld~\cite{Rosenfeld:1940}
and  Belinfante~\cite{Belinfante:1940} (and as we will discuss in subsection~\ref{sec:EHtensor}),
this conceptually quite different approach yields, in the flat space
 limit of the minimal coupling to gravity,
Belinfante's symmetric \EMT whose construction does not make any reference to gravity.
(In particular one recovers the canonical \EMT if this one is already symmetric.
We note that Belinfante's symmetric \EMT admits a natural geometric formulation
if gravity is viewed as a gauge theory of the Poincar\'e group, see
references~\cite{Blagojevic:2013xpa, Grensing:2013}.)
One can also recover the traceless Callan-Coleman-Jackiw EMT  in the flat space limit of a conformally invariant
field theory on curved space.

\paragraph{Summary:}
\emph{If one takes for granted that Einstein's general relativity is the correct theory for the gravitational field,
then the \EMT
of gauge and/or matter fields
in Minkowski space should not only be conserved and gauge invariant,
but it should also be symmetric and naturally generalizable to a generally covariant tensor in curved space.}
Yet, we should note  that within certain alternative theories of gravity
like Einstein-Cartan
theory~\cite{\hehlhammond,TrautmanEMP, Blagojevic:2013xpa},
the canonical \EMT appears naturally.
In particular, non symmetric EMT's which do exist for certain classical spin fluids can consistently be coupled to gravity
in this framework, see~\cite{Hehl:2014eja} and references therein.
Concerning the validity of the different theories of gravity we note that Einstein-Cartan theory is considered
to be a viable alternative to general relativity: both theories can only be distinguished at very high densities or at very small
distances which are currently beyond experimental reach --- see~\cite{Hehl:2014eja} and references therein.

The canonical and the improved (or Einstein-Hilbert-) EMT's only differ by a total derivative so that they yield the \emph{same conserved charges $P^\n$}
and it is the latter which play a fundamental role in classical and quantum field theories on flat
space~\cite{Hanson:1976cn, IP}:
They appear for instance
in the defining axioms of Wightman and of Haag and Kastler for quantum field
theory~\cite{Streater:1989vi, Haag:1992hx}.

\subsection{Improvement of the canonical EMT's in gauge theories }\label{sec:diffimpEMT}

\subsubsection{Belinfante's procedure for constructing a symmetric EMT}\label{sec:BelProc}

\paragraph{General procedure:}
The concern of Belinfante~\cite{Belinfante:1939, Belinfante:1940} was to construct an improved EMT which is
symmetric  in its indices for any relativistically invariant field theory.
The resulting tensor $T_{\txt{imp}}^{\mu \nu}$ which is also referred to as the
\emph{Belinfante(-Rosenfeld) tensor,}
is obtained by choosing the superpotential
appearing in equation~\eqref{eq:imprEMT}
to be given by
\begin{align}
\chi^{\mu \rho \nu} = - \frac{1}{2} \, (  s^{\mu \rho \nu} - s^{\nu \mu \rho } + s^{\rho \nu \mu})
\, ,
\label{eq:BelinfantePot}
\end{align}
where $s^{\mu \rho\nu}$ represents the
spin density tensor of the fields which appears in the canonical angular momentum tensor \eqref{eq:canAMT}.
For a detailed presentation of this approach we refer to the literature, e.g.
see reference~\cite{Grensing:2013}. Here we
emphasize that this approach applies to any relativistically invariant field theory, but that,
for gauge theories,
the gauge invariance of the ``symmetry improved'' EMT of Belinfante is not a priori ensured.
We will come back to this point in footnotes \ref{fn4} and \ref{fn5}.

\paragraph{Pure gauge fields:}
For the YM field, the described improvement procedure
for the canonical EMT~\eqref{eq:canEMTYM}
yields $\chi^{\m \rho \n} = F_a^{\m \rho} A_a^{ \n}$.
Use of the equation of motion $D_\rho F^{\m \rho}=0$
then leads to  the tensor
\begin{align}
\Boxed{
T_{\txt{imp}}^{\mu \nu} (F) =
 \frac1{c_2} \, \Tr \, ( F^{\mu \rho}   {F_\rho}^{\nu} + \frac{1}{4} \, \eta^{\mu \nu} F^{\rho \sigma}  F_{\rho \sigma} )
 }
\, ,
\label{eq:impEMTYM}
\end{align}
which is not only conserved and symmetric, but also gauge invariant\footnote{\label{fn4}%
As a matter of fact,
the canonical expression~\eqref{eq:canEMTYM}  can be rendered gauge invariant in a very direct way
by guessing the superpotential term $\pa_{\rho} (F_a^{\mu \rho} A_a^\nu )$ since this term reduces on-shell (i.e. by virtue
of the equation of motion $D_\rho F^{\m \rho } =0$)
to $F_a^{\mu \rho} (\pa_{\rho} A^\nu + \ri g \, [A_\rho , A^\n ])_a$ and therefore allows to ``gauge covariantize''
the factor $\pa^\nu A_\rho$ in $T_{\txt{can}}^{\mu \nu} $,
thus ensuring gauge invariance of the EMT.
This way of proceeding
 directly yields the result~\eqref{eq:impEMTYM} without determining the spin density $s^{\m \rho \n}$.}.
In $n=4$ dimensions it is traceless.

Let us briefly consider the particular case of Maxwell's $U(1)$ theory in four space-time dimensions, for which the tensor~\eqref{eq:impEMTYM}
reduces to $T_{\txt{imp}}^{\mu \nu} (F) =F^{\mu \rho}   {F_\rho}^{\nu} + \frac{1}{4} \, \eta^{\mu \nu} F^{\rho \sigma}  F_{\rho \sigma} $.
This tensor encodes the familiar expressions~\cite{Jackson:1998} for the energy and momentum densities
of the electromagnetic field (i.e. $T_{\txt{imp}}^{00} = \frac{1}{2} \, (\vec E ^{\; 2} + \vec B ^{\; 2} )$ and
$T_{\txt{imp}}^{0i} = (\vec E \times \vec B )_{x^i}$)
as well as for Maxwell's stress tensor density (given by the spatial components $T_{\txt{imp}}^{ij}$).
\emph{The result for $T_{\txt{imp}}^{\mu \nu} (F)$
 expresses in a remarkable manner the union of space and time, energy and momentum
as well as electricity and magnetism} as achieved by the masters of electrodynamics and special relativity (Maxwell,
Einstein, Lorentz, Poincar\'e and Minkowski).
For $\n =0$, the local conservation law $\pa_\m T_{\txt{imp}}^{\mu \nu}  =0$ is simply \emph{Poynting's theorem} for the free electromagnetic field.
Furthermore, we have $T_{\txt{imp}}^{00} (F) \geq 0$
(and $  T_{\txt{imp}}^{00} (F) =0$ if and only if $F=0$),
as well as $| \vec E \times \vec B \, | \leq  \frac{1}{2} \, (\vec E ^{\; 2} + \vec B ^{\; 2} ) = T_{\txt{imp}} ^{00}$:
from these relations it follows that $|\vec P \, | \leq P^0$ for the
components $P^\n \equiv \int_{\br ^3} d^3x \, T_{\txt{imp}}^{0\n}$ of the
energy-momentum four-vector,
i.e. the latter vector is time-like and future directed~\cite{Thirring:1997}.

\paragraph{Matter fields interacting with a gauge field:}
The improvement procedure of Belinfante is somewhat different and more subtle
when matter fields (scalars or fermions) are coupled to a gauge field
(e.g. see reference~\cite{Grensing:2013} for Abelian gauge theory).
By way of illustration, let us consider a multiplet $\psi \equiv [\psi_1 , \dots, \psi_N ]^t$ of Dirac fields
which is minimally coupled
to the YM field, i.e. we have the Lagrangian density~\eqref{eq:LMDiracYM}:
${\cal L}_M (\psi, A) \equiv {\ri} \, \bar{\psi} \gamma^\m \! \stackrel{\leftrightarrow}{D}_{\mu} \!\! \psi - m \bar{\psi} {\psi}$.
Then, the canonical EMT reads
\begin{align}
T_{\txt{can}}^{\mu \nu} ( \psi, A ) = \frac{\ri}{2} \, \Big[ \bar{\psi} \gamma^\m {D}^{\nu} \psi -  ({D}^{\nu} \bar{\psi} ) \gamma^\m \psi  \Big]
 - \eta^{\m \n} {\cal L}_M + A^\n _a j_a^\m [ \psi, A ]
 \, ,
 \label{eq:CanEMTpsi}
\end{align}
where $j_a^\m [ \psi, A ] \equiv \mq \bar{\psi} \gamma^\m T_a \psi$.
In this case,
the standard (``naive'') improvement procedure is not sufficient,  rather \emph{an extra minimal coupling procedure has to be performed
at the level of the EMT's} so as to discard the last term (i.e. the potential-current term) in expression~\eqref{eq:CanEMTpsi}.
We will see in the next section that the improved EMT's follow straightforwardly
from the canonical EMT's if we improve these tensors by taking into account
local gauge invariance.

\subsubsection{Improvement procedure for constructing a gauge invariant EMT}
\label{sec:impgaugetheory}

For pure gauge theories (Abelian or non-Abelian) the standard derivation
of the EMT
(which consists of determining the consequence of the invariance of the action functional under space-time translations as outlined
at the beginning of~\secref{sec:IntrodEMT})
 does not take into account the gauge invariance of the Lagrangian.
Thus, the lack of gauge invariance of the canonical EMT does not come as a surprise.
\emph{Gauge invariance of the EMT}
can either be achieved by exploiting this invariance of the Lagrangian
in the course of the derivation of the EMT or by exploiting it upon improving the canonical EMT.
The first approach is discussed in detail in reference~\cite{Munoz:1996wp}
(see also~\cite{Eriksen:1979vq,  Takahashi:1985dt, Felsager:1981iy,\montesinos})
and we will present here the second procedure which is inspired by the first one, but which has not been considered
in the literature to the best of our knowledge.
As we will see, the latter approach can also be used
to determine in a neat and straightforward manner the improved EMT for massive Abelian vector fields, i.e. Proca theory
(this result being inspired by the work~\cite{Felsager:1981iy, \montesinos}) despite the fact that the complete Lagrangian
is not gauge invariant in this case.
For all of these theories,
the improved EMT obtained by exploiting gauge invariance
(considering either of the two methods that we just outlined)
 coincides with the one obtained
by Belinfante's procedure whose goal is to construct a symmetric EMT out of the canonical one.
More generally, we will show that our
 method of exploiting gauge invariance also generalizes to the case where matter fields
(scalar or spinor fields) are coupled in a gauge invariant manner to gauge fields: in this context the improvement follows readily
and there is no need to add by hand some extra term as it is the case for Belinfante's procedure applied
to this physical system.
We will consider a general gauge theory, the case of an Abelian theory
corresponding to the choice $a=1$ for the internal symmetry indices and  $f^{abc} =0$ for the structure constants.

\paragraph{Improvement for pure gauge theories:}
The assumptions concerning the Lagrangian ${\cal L}$ are as follows.
As before, we assume that ${\cal L} (x) \equiv {\cal L}( A_\nu (x), \pa_\mu A_\nu (x))$ is a scalar field depending on $A_\nu$
and its first-order derivatives, but not explicitly on $x$. Moreover, we suppose
 that ${\cal L}$ has the form
\begin{align}
\label{eq:gaugeLag}
{\cal L} \equiv  {\cal L}_g (F) + {\cal L}_m (A)
 \, ,
 \end{align}
 where $ {\cal L}_m (A)$ only depends on $A_\n$ (and not on its derivatives) and
where  ${\cal L}_g$ is assumed to be
 gauge invariant. The latter assumption implies that ${\cal L}_g$ only depends on $A_\nu$ and its derivatives  by virtue of the
components $F_{\mu \nu}$ of the field strength tensor,
this tensor being given by expression~\eqref{eq:YMfs}, i.e.
${F} _{\mu \nu} = \pa_{\mu} {A} _{\nu} - \pa_{\nu}
{A} _{\mu} + \ri \mq \, [ {A} _{\mu}  ,  {A} _{\nu} ]$.
(This result follows for instance explicitly
by exploiting the so-called Klein-Noether identities~\cite{Sundermeyer:2014kha}
which have  first been discussed by F.~Klein for general relativity and which have been rediscovered by R.~Utiyama~\cite{Utiyama:1959}
in the context of YM theories.)
Henceforth we have
\begin{align}
\label{eq:derivLag}
\frac{\pa {\cal L}}{\pa(\pa_\mu A_\nu ^a)} =
\frac{\pa F_{\rho \sigma} ^b}{\pa(\pa_\mu A_\nu ^a)} \,
\frac{\pa {\cal L}_g}{\pa F_{\rho \sigma} ^b}
= 2 \, \frac{\pa {\cal L}_g}{\pa F_{\mu \nu} ^a}
 \, .
 \end{align}

By definition
the canonical EMT~\eqref{eq:canEMT} presently reads
$$
T_{\txt{can}}^{\mu \nu} = \frac{\pa {\cal L}}{\pa(\pa_\mu A_\rho ^a)} \, \pa^{\nu}  A_\rho ^a  - \eta^{\mu \nu} {\cal L}
\, ,
$$
and obviously fails to be gauge invariant (even for ${\cal L}_m =0$)
due to the factor $\pa^{\nu}  A_\rho ^a $ in its first term.
To obtain a gauge invariant expression, we use relations~\eqref{eq:derivLag}
and~\eqref{eq:YMfs}:
\begin{align}
T_{\txt{can}}^{\mu \nu} \equiv
2 \, \frac{\pa {\cal L}_g}{\pa F_{\mu \rho} ^a}
\left( F^{\nu}_{a \, \rho} + \pa_{\rho} A^\nu _a - \ri \mq \, [ A^\nu , A_{\rho} ] _a \right)
 - \eta^{\mu \nu} {\cal L}
\, .
\label{eq:canEMTgrev}
\end{align}
For the second term we now apply the Leibniz rule:
\begin{align}
2 \, \frac{\pa {\cal L}_g}{\pa F_{\mu \rho} ^a} \, \pa_{\rho} A^\nu _a =
- \pa_{\rho} \chi^{\mu \rho \nu} -  2 \pa_\rho
\Big( \frac{\pa {\cal L}_g}{\pa F_{\mu \rho} ^a} \Big) A_a^\nu
\, , \qquad \txt{with} \ \;
 \chi^{\mu \rho \nu} \equiv
- 2 \, \frac{\pa {\cal L}_g}{\pa F_{\mu \rho} ^a} A_a^{\nu} = -  \chi^{\rho \mu \nu}
\, .
\label{eq:RewriteSup}
\end{align}
Here, the contribution $\pa_{\rho} \chi^{\mu \rho \nu}$ is a superpotential term\footnote{\label{fn5}%
We note
that for ${\cal L}_g  \equiv - \frac{1}{4} \, F_a^{\mu \nu} F^a_{\mu \nu} $,
we have
$ \chi^{\mu \rho \nu} = F^{\mu \rho}_a A^\n _a$: This superpotential
coincides with the one resulting from Belinfante's improvement procedure applied to pure gauge theories.}
so that we have obtained an improved EMT $T_{\txt{imp}}^{\mu \nu} $
 of the form~\eqref{eq:imprEMT}.
The second term on the right hand side of the
previous equation can be rewritten by using~\eqref{eq:derivLag}
and the equation of motion of the gauge field,
$ \pa_\rho
(\frac{\pa {\cal L}}{\pa (\pa _\rho A_{\mu} ^a )} ) = \frac{\pa {\cal L}}{\pa  A_{\mu} ^a }$:
\begin{align}
-  2 \pa_\rho \Big( \frac{\pa {\cal L}_g}{\pa F_{\mu \rho} ^a} \Big) A_a^\nu =
\pa_\rho \Big( \frac{\pa {\cal L}}{\pa (\pa _\rho A_{\mu} ^a )} \Big) A_a^\nu
= \frac{\pa {\cal L}}{\pa  A_{\mu} ^a } \, A_a^\nu
\, .
\label{eq:intermed}
\end{align}
For the Lagrangian~\eqref{eq:gaugeLag} we have
$\frac{\pa {\cal L}}{\pa  A_{\mu} ^a } = \frac{\pa {\cal L}_g (F) }{\pa  A_{\mu} ^a } +
\frac{\pa {\cal L}_m (A)}{\pa  A_{\mu} ^a }$
with
\begin{align}
\frac{\pa {\cal L}_g  }{\pa  A_{\mu} ^a } = \frac{\pa F_{\rho \sigma}^b}{\pa  A_{\mu} ^a }
\frac{\pa {\cal L}_g }{\pa F_{\rho \sigma}^b }
= -2\mq  f^{bca} A_\rho ^c \,
\frac{\pa {\cal L}_g  }{\pa F_{\rho \m}^b }
\, ,
\label{eq:DerLA}
\end{align}
and thereby~\eqref{eq:intermed} becomes
\begin{align}
-  2 \pa_\rho \Big( \frac{\pa {\cal L}}{\pa F_{\mu \rho} ^a} \Big) A_a^\nu
=
2\ri \mq \, [A^\nu , A_\rho ]_a \, \frac{\pa {\cal L}_g  }{F^a_{\mu \rho} }
+ \frac{\pa {\cal L}_m}{\pa  A^a_{\mu} } \, A_a^\nu
\, .
\label{eq:RewrDL}
\end{align}
Since the first
 term in this expression compensates the third term in expression~\eqref{eq:canEMTgrev}, we are led
to the final result $T_{\txt{imp}}^{\mu \nu} = T_{\txt{can}}^{\mu \nu} +  \pa_{\rho} \chi^{\mu \rho \nu} $ with
\begin{align}
\Boxed{
T_{\txt{imp}}^{\mu \nu} =
T_{\txt{imp}}^{\mu \nu} (F) + T_{\txt{imp}}^{\mu \nu} (A)
\equiv
\Big[ 2 \, \frac{\pa {\cal L}_g }{\pa F_{\mu \rho}^a } \, {F^\nu_a}_\rho
- \eta^{\mu \nu} {\cal L}_g \Big]
+ \Big[
\frac{\pa {\cal L}_m}{\pa  A_{\mu}^a  } \, A^\nu_a
- \eta^{\mu \nu} {\cal L}_m \Big]
}
\, .
\label{eq:impEMTgauge}
\end{align}
For the Lagrangian ${\cal L} \equiv {\cal L}_g (F)$, this EMT only
depends on the gauge potential by means of the field strength, hence
its gauge invariance is ensured by the invariance of ${\cal L}_g (F)$.

As a first application we consider \textbf{pure YM theory}, i.e. ${\cal L} = {\cal L}_g (F) \equiv
- \frac{1}{4c_2} \, \Tr \, (F^{\mu \nu} F_{\mu \nu} )$,
thus we recover the result~\eqref{eq:impEMTYM}.
By construction, this EMT is conserved and gauge invariant.
It is also symmetric
and it is traceless for $n=4$; the result coincides with the one obtained by Belinfante's procedure
(see footnote \ref{fn5}) or from the Einstein{-}Hilbert EMT (\secref{sec:gravity}).

If we include a mass term for the vector field $A_{\mu}$ into the
four dimensional Abelian theory by adding
${\cal L}_m (A) \equiv \frac{1}{2} \, m^2 A^\mu A_\mu$ to ${\cal L}_g (F)$, then~\eqref{eq:impEMTgauge} yields
\begin{align}
 T_{\txt{imp}}^{\mu \nu} =  T_{\txt{imp}}^{\mu \nu} (F) + m^2 (A^\mu A^\nu - \frac{1}{2} \, \eta^{\mu \nu} A^\rho A_\rho )
\, .
\end{align}
This is the improved EMT for the \textbf{Proca theory}~\cite{Felsager:1981iy}
where the mass term is also symmetric, but neither gauge nor scale invariant.

\paragraph{Improvement for matter fields interacting with gauge fields:}

Consider charged matter fields  $\vp$ (complex scalars $\phi$ or Dirac spinors $\psi$)
which are minimally coupled to gauge fields.
For notational clarity, we first discuss the case of a multiplet of \textbf{complex scalar fields}, i.e.
$\vp = (\phi, \phi^\dagger)$, for which we have specified the dynamics in equations~\eqref{eq:LagGM}-\eqref{eq:CovConsLaw}.
Quite generally, suppose that we have a \emph{gauge invariant} Lagrangian of the form
\[
{\cal L} \equiv {\cal L}_g (F)   + {\cal L}_M \, ,
\quad \txt{with} \ \;
{\cal L}_M  \equiv {\cal L}_0 (D_\m \vp ) +  {\cal L}_1 (\vp )
\, ,
\]
where $ {\cal L}_1 (\vp )$ only depends on the field $\vp$ and not on its derivatives.
By virtue of gauge invariance,
 the kinetic terms for gauge and matter fields, i.e. ${\cal L}_g (F)$ and $ {\cal L}_0 (D_\m \vp )$
can only depend on the field strength and covariant derivatives, respectively:
\begin{align}
F_{\m \n } \equiv \pa_\m A_\n - \pa_\n A_\m  + \ri \mq \, [ {A} _{\mu}  ,  {A} _{\nu} ]
\, , \qquad
D_\m \phi \equiv \pa_\m \phi + \ri \mq {A}_\m \phi
\, , \qquad
D_\m \phi^\dagger \equiv  \pa_\m \phi^\dagger - \ri \mq \phi^\dagger {A}_\m
\, .
\label{eq:Feldstaerke}
\end{align}
Here, the generators $T_a$ of the underlying Lie algebra are assumed to be in the adjoint representation
for $F_{\m \n}$ and in an $N$-dimensional representation for the multiplet of matter fields.

As noted in equation~\eqref{eq:genNAmattercurrent}, the variation of the matter field action
$S_M [\vp, A] \equiv \int d^nx \, {\cal L}_M $ with respect to the gauge field yields the matter current:
\begin{align}
j_a^\mu [\vp , A ] \equiv - \frac{\delta S_M [\vp, A]}{\delta A^a _\mu}
= - \frac{\pa {\cal L}_M}{\pa A^a _\m} =  \ri \mq \Big[  \phi^\dagger {T}_a \, \frac{\pa {\cal L}_0}{\pa (D_\mu \phi^\dagger)}
-  \frac{\pa {\cal L}_0}{\pa (D_\mu \phi)} \,  {T}_a \phi \Big]
\, .
\label{eq:Stromdichte}
\end{align}

By its very definition~\eqref{eq:canEMT},
the canonical EMT of the physical system under consideration is given by
\begin{align}
T_{\txt{can}}^{\mu \nu} = \frac{\pa {\cal L}}{\pa (\pa_{\mu} A^a _\rho )} \, \pa^{\nu} A^a _\rho
+\frac{\pa {\cal L}}{\pa (\pa_{\mu} \phi )} \, \pa^{\nu} \phi
+ (\pa^{\nu} \phi^\dagger) \,  \frac{\pa {\cal L}}{\pa (\pa_{\mu} \phi^\dagger )}
- \eta^{\mu \nu} {\cal L}
\, .
\label{eq:KanEIT}
\end{align}
Let us now substitute relation~\eqref{eq:derivLag} and $\frac{\pa {\cal L}}{\pa (\pa_{\mu} \vp )} = \frac{\pa {\cal L}}{\pa (D_{\mu} \vp )}$,
and let us use~\eqref{eq:Feldstaerke} to express the ordinary derivatives $ \pa^{\nu} A^a _\rho$ and $\pa^\n \vp$ appearing in~\eqref{eq:KanEIT} in terms of
the field strength and the covariant derivatives, respectively.
After grouping together the different contributions and substituting relation~\eqref{eq:Stromdichte}, we get the result
\begin{align}
T_{\txt{can}}^{\mu \nu} = T_{\txt{int}}^{\mu \nu} (F) + T_{\txt{int}}^{\mu \nu} (\vp, A) + A_a ^\n j_a ^\mu [\vp , A ]
+ 2 \, \frac{\pa {\cal L}_g }{\pa F^a _{\m \rho}} \, \Big( \pa_\rho A_a ^\n + \ri \mq \, [ A_\rho , A^\n ]_a \Big)
\, ,
\label{eq:TotKanEIT}
\end{align}
with
\begin{align}
\label{eq:IntEIT}
\Boxed{
T_{\txt{int}}^{\mu \nu} (F) \,  \equiv 2 \, \frac{\pa {\cal L}_g}{\pa F^a _{\m \rho}} \, {F_a ^\n}_\rho - \eta^{\m \n} {\cal L}_g
= T_{\txt{imp}}^{\mu \nu} (F)
}
\end{align}
and
\begin{align}
\label{eq:IntEITmatter}
\Boxed{
T_{\txt{int}}^{\mu \nu} (\vp, A) \,  \equiv \frac{\pa {\cal L}_M}{\pa (D_{\mu} \phi )} \, D^\n \phi
+ ( D^\n \phi^\dagger ) \, \frac{\pa {\cal L}_M}{\pa (D_{\mu} \phi^\dagger )} - \eta^{\m \n} {\cal L}_M
}
\, .
\end{align}
To conclude, we rewrite the last term in~\eqref{eq:TotKanEIT}
following the line of arguments~\eqref{eq:RewriteSup}-\eqref{eq:RewrDL}:
\[
2 \, \frac{\pa {\cal L}_g }{\pa F^a _{\mu \rho}} \, \pa_{\rho} A_a ^\nu
=- \pa_{\rho} \chi^{\mu \rho \nu} + \frac{\pa {\cal L}}{\pa  A^a_{\mu} } A^\nu _a
= - \pa_{\rho} \chi^{\mu \rho \nu}
- 2\ri \mq \, \frac{\pa {\cal L}_g }{\pa F^a _{\m \rho}} \, [ A_\rho , A^\n ]_a
-  A_a ^\n j_a ^\mu [\vp , A ]
\, ,
\]
where the last expression follows from  $\frac{\pa {\cal L}}{\pa  A^a _{\mu} } =
\frac{\pa {\cal L}_g}{\pa  A^a _{\mu} } + \frac{\pa {\cal L}_M}{\pa  A^a _{\mu} }$
and the definition~\eqref{eq:Stromdichte}.

In summary, with the definition
\[
T_{\txt{int}}^{\mu \nu} \equiv T_{\txt{can}}^{\mu \nu} +  \pa_{\rho} \chi^{\mu \rho \nu} \, , \qquad
\txt{where} \ \; \chi^{\mu \rho \nu} \equiv - 2\, \frac{\pa {\cal L}_g}{\pa F^a _{\mu \rho}} \,   A_a ^\n
\, ,
\]
we have a \emph{total EMT $T_{\txt{int}}^{\mu \nu}$ for the interacting system of fields} which is a sum
of the improved EMT for the gauge field and the EMT of the matter fields interacting with the gauge field:
\begin{align}
\Boxed{
T_{\txt{int}}^{\mu \nu} = T_{\txt{int}}^{\mu \nu} (F) + T_{\txt{int}}^{\mu \nu} (\vp, A)
 }
\, .
\label{eq:TotintEIT}
\end{align}
Upon considering the Lagrangians~\eqref{eq:invact}, \eqref{eq:LagGM}, i.e.
${\cal L}_g (F) \equiv  \frac{-1}{4 c_2} \,  \Tr
\, ( {F} ^{\mu \nu} {F} _{\mu \nu} )$ and
${\cal L}_M (\phi , A) \equiv (D^\mu \phi^\dagger) ( D_\mu \phi ) - m^2 \phi^\dagger \phi$,
we obtain the explicit expression~\eqref{eq:impEMTYM} and
\begin{align}
\label{eq:IntscalEIT}
\Boxed{
T_{\txt{int}}^{\mu \nu} (\phi ,A)  = (D^{\mu} \phi^\dagger ) ( D^\n \phi)
+ (D^{\nu} \phi^\dagger ) ( D^\mu \phi) - \eta^{\m \n} {\cal L}_M (\phi ,A)
 }
\, .
\end{align}
Each of the tensors~\eqref{eq:impEMTYM}, \eqref{eq:IntscalEIT}
is {gauge invariant} (by construction)
and {symmetric} in its indices. The total EMT  $T_{\txt{int}}^{\mu \nu}$
is conserved (by construction), but the different contributions are not:
their divergence
can easily be evaluated by using the field equations~\eqref{eq:EOMmatterMul} and~\eqref{eq:YMeq}
as well as the commutation relations $[D_\m , D_\n ] \phi = \ri \mq F_{\m \n} \phi$
and $[D_\m , D_\n ] \phi^\dagger = - \ri \mq \phi^\dagger F_{\m \n} $:
\begin{align}
\label{eq:divEMTs}
\Boxed{
\pa_\m T_{\txt{int}}^{\mu \nu} (F) =  \frac{1}{ c_2} \, \Tr
\, ( j_\mu F^{\m \n})
= - \pa_\m T_{\txt{int}}^{\mu \nu} (\phi, A )
 }
\, , \qquad \txt{hence} \quad \
\Boxed{
\pa_\m T_{\txt{int}}^{\mu \nu} =0
 }
\, .
\end{align}
Thus, we have the standard  \emph{local conservation law for the total EMT tensor of the interacting system.}
We note that the partial differential equation $\pa_\m T_{\txt{int}}^{\mu \nu} (\phi , A) = -  \frac{1}{ c_2} \, \Tr
\, ( j_\mu  F^{\m \n}) $
can be viewed as a \emph{continuum version of the Lorentz-Yang-Mills force law}, i.e. Wong's equation~\cite{Balasin:2014dma}.
The relation $\pa_\m T_{\txt{int}}^{\mu \nu} (F) =  \frac{1}{ c_2} \, \Tr
\, (j_\mu F^{\m \n})$ may be regarded
 as the balance equation for the gauge field energy.
Indeed (considering Maxwell's  $U(1)$ theory), from
$(j ^\m ) \equiv (\rho , \vec{j} \, )$ we have the Lorentz force density
$( F^{\m \n} j_\nu) = (\vec j \cdot \vec E , \rho \vec E + \vec{j} \times \vec B \, )$, hence  we obtain, for $\n =0$, the local balance equation
\[
- \pa_t w = \Div\, \vec{ {\cal P} } + \vec{j} \cdot \vec E
\, ,
\]
where $ w \equiv \frac{1}{2} \, (\vec E^{\, 2} + \vec B ^{\, 2} \,)$ represents the electromagnetic energy density
and $\vec{ {\cal P} } \equiv \vec E \times \vec B$
is the Poynting vector.
The latter result is nothing else but \emph{Poynting's theorem} expressing that the electromagnetic fields not only radiate, but also do work on the electric charges
(currents), thereby transforming electromagnetic energy into mechanical or thermal energy.

The case of a \textbf{Dirac field} is tantamount to considering $\vp = (\psi_{\alpha}, \bar{\psi}_{\alpha})$
in the previous derivation. The expression for $T_{\txt{int}}^{\mu \nu} (F)$ is unchanged and for
$T_{\txt{int}}^{\mu \nu} (\psi, A )$ one obtains the following explicit result
from the Lagrangian ${\cal L}_M (\psi) \equiv {\ri} \, \bar{\psi} \gamma^\m \! \stackrel{\leftrightarrow}{D}_{\mu} \!\! \psi
- m \bar{\psi} \psi$:
\begin{align}
\label{eq:IntDiracEIT}
T_{\txt{int}}^{\mu \nu} ( \psi, A ) = \frac{\ri}{2} \, \Big[ \bar{\psi} \gamma^\m D^{\nu} \psi
- (D^{\nu} \bar{\psi} ) \gamma^\m \psi  \Big] - \eta^{\m \n} {\cal L}_M
\, .
\end{align}
By construction, this expression is \emph{gauge invariant}, but it is not symmetric in its indices.
The symmetry can be achieved by hand
($T_{\txt{int}}^{( \mu \nu )} \equiv \frac{1}{2} \, [  T_{\txt{int}}^{\mu \nu} + T_{\txt{int}}^{\nu \mu} ]$)
without destroying the other characteristics of the tensor:
\begin{align}
\label{eq:IntDiracEMT}
\Boxed{
T_{\txt{int}}^{(\mu \nu)} (\psi, A)   = \frac{\ri}{4} \, \Big[ \bar{\psi} \gamma^\m D^{\nu} \psi
- (D^{\nu} \bar{\psi} ) \gamma^\m \psi + (\mu \leftrightarrow \nu )  \Big] - \eta^{\m \n} {\cal L}_M
}
\, .
\end{align}

\paragraph{Summary:}
We have recovered in a constructive manner the same results as for the modified Belinfante procedure.
The present approach does not require a determination of the spin densities and of the associated superpotential terms, nor
does it require to perform specific minimal couplings for each of the tensors under consideration.
The only shortcoming is that the symmetry of the ``gauge improved'' EMT is not automatically realized.
In fact, the symmetrization of the gauge improved EMT for the Dirac field has to be achieved by hand, but this does not raise
any problem and this feature is actually also encountered for the Einstein{-}Hilbert EMT of the Dirac field,
see equation~\eqref{eq:symspinEMT} below.

\subsection{Assessment concerning the improvements}

As we discussed in equations~\eqref{eq:AMT}-\eqref{eq:canAMT}, the symmetry of the EMT in Minkowski space is convenient,
but by no means mandatory.
Moreover it does not become mandatory when matter or gauge fields are coupled to gravity
if one takes into account the fact that the currently available experimental
data do not allow us to discriminate between general relativity and alternative theories of Einstein-Cartan-type which allow for torsion
and for a non symmetric EMT of matter or gauge fields.

If one considers gauge field theories in Minkowski space as we do in the present article,
then the EMT necessarily has to be gauge invariant due to its physical interpretation.
However, Belinfante's improvement procedure does not yield \emph{a priori} a gauge invariant EMT
when applied to gauge theories, and in addition it does not work in the
straightforward manner for the physically interesting case where
matter fields are minimally coupled to a gauge field. The improvement procedure that we presented here
is devised  to
obtain an EMT which satisfies the physically unavoidable condition of gauge invariance and it readily works
for pure gauge theories as well as for matter fields interacting with gauge fields
(and even for massive Abelian gauge fields).
For pure gauge fields, the resulting expression for the ``gauge-improved'' EMT coincides with the EMT obtained by Belinfante's
``symmetrization procedure''
since the involved superpotential terms coincide with each other.

\section{Einstein-Hilbert EMT in Minkowski space}\label{sec:gravity}

\subsection{Motivation and procedure}
The EMT $(T^{\m \n})$ in Minkowski space
represents a collection of conserved current densities which are associated to
space-time translations, i.e. geometric transformations which act on both space-time coordinates and on fields.
These transformations being more complex than internal symmetries
which only act in the space of fields,
it is useful to look at the conserved current densities associated to \emph{internal symmetries} so as to address
the conceptual and technical issues in a simpler setting.

The Lagrangian ${\cal L}_M (\vp) \equiv {\cal L}_M (\vp , A=0)$ for \emph{free} charged matter fields
is invariant under \emph{global} (rigid) gauge transformations, i.e. at the infinitesimal level under
transformations parametrized by real constants $\omega^a$ with $a \in \{ 1, \dots, n_G\}$.
(More generally, one can consider a gauge invariant self-interaction of matter fields,
e.g. include a self-interaction potential $V(\phi^{\dagger} \phi)$ for a multiplet of scalar fields
(like the Higgs field) or an invariant Yukawa-type coupling between scalar and spinor fields.
Since these interaction terms do not involve derivatives of fields, they lead to the same canonical currents~\eqref{eq:CanCurrMatt}.)
According to Noether's first theorem, the canonical currents $j^\m_{\txt{can},a} [\vp ]$
associated to the global gauge invariance are given by
$ \omega ^a j^\m _{\txt{can},a} [\vp] = \frac{\pa {\cal L}_M (\vp)}{\pa (\pa_\m \vp )} \, \delta \vp$,
where $\delta \vp$ denotes an infinitesimal global gauge transformation.
For multiplets of free complex scalars or Dirac spinors, one has the respective expressions
\begin{align}
\label{eq:CanCurrMatt}
j ^\m _{\txt{can},a} [\phi] = \ri \mq \Big[  \phi^\dagger {T}_a \pa^\mu \phi
-  (\pa^{\mu } \phi^\dagger) {T}_a  \phi \Big]
\, ,
\qquad
j_{\txt{can},a} ^\m [\psi] = \mq \, \bar{\psi}  \gamma^{\mu } {T}_a \psi
\, .
\end{align}
Here, the coupling constant $\mq$ represents the ``non-Abelian'' charge of fields
which we have factored out from the parameters $\omega^a$ so as to have the same overall
numerical factor for the canonical Noether currents as for the currents resulting from a coupling to a gauge field, see next equation.

Equivalently, the expressions~\eqref{eq:CanCurrMatt} can be viewed as the response
of the physical system to a variation of the (external) gauge field,
the latter field mediating the interaction between charge carrying matter fields
(see equations~\eqref{eq:genNAmattercurrent}, \eqref{eq:matterNAcurrent}):
 \begin{align}
\Boxed{
j_{\txt{can},a}^\mu [\vp] = -
\left. \frac{\delta S[\vp ;A]}{\delta A^a _\mu} \right| _{A=0}
}
\qquad (A = \, \mbox{external gauge field})
\,.
\label{eq:relatecurrents}
\end{align}
The equivalence follows from the fact that for an external gauge field, the action $S[\vp ;A]$
is in particular invariant under global gauge transformations and reduces to the free
field action for $A=0$.

In this section, we discuss the generalization of these ideas to geometric symmetries
(the global differential geometric aspects having been elaborated in detail in reference~\cite{Forger:2003ut}).
More precisely,
we apply the lines of reasoning described above for internal symmetries to
 the case of the currents $T^{\m \n} _{\txt{can}}$ associated to the invariance
of the relativistic field theory in
Minkowski space under space-time translations, i.e. under the rigid geometric
transformations $x \leadsto x'(x) = x+a$ with $a^\m$ constant. In this respect
we first note that the associated conserved Noether charges $P^\n = \int d^{n-1} x \, T^{0\n}_{\txt{can}}$
represent the energy-momentum of the fields, these fields being matter fields (describing for instance electrons)
or gauge fields (gauge bosons mediating the interaction).
The exchange of energy and momentum is realized by gravity as described by a metric,
i.e. a symmetric non-degenerate
 tensor field  $\mg (x) \equiv \left( g_{\mu \nu}  (x) \right)$.
Both matter and gauge fields (or particles representing their field quanta) are subject to the gravitational interaction.

In section~\ref{sec:fieldsCurvedST}, we first consider the case of a \emph{dynamical} gravitational field,
the gravitational analogue of the YM equations $D_\m F^{\m \n } = j^\n$ for the gauge field being given by
Einstein's field equations $G_{\m \n} = - \kappa T_{\m \n}$ for the gravitational field: instead of gauge transformations
(i.e. local internal transformations) we now consider diffeomorphisms $x^\mu \leadsto x'^\m (x) \simeq
x^\m + \xi^\m (x)$
(i.e. general coordinate transformations or $x$-dependent translations).
By definition, the EMT $T^{\m \n} [\vp , \mg]$
in curved space then represents the response of the physical system to a variation of the
(external) gravitational field $g_{\m \n}$, and its covariant conservation law in curved space holds by virtue of the reparametrization invariance
of the action.
Finally, in section~\ref{sec:EHtensor} we consider the flat (Minkowski) space limit $g_{\m \n} = \eta_{\m \n}$ (representing
the analogue of $A=0$ for the internal symmetries) and we argue that the resulting EMT
$T^{\m \n} [\vp , \mg = \eta]$
(to which we refer to as the Einstein-Hilbert EMT)
coincides with the improved  EMT
$T^{\m \n}_{\txt{imp}} [\vp ]$ associated to translational invariance
in Minkowski space. For this identification an ambiguity appears since a given Minkowski space
Lagrangian
may represent the flat space limit of different non-equivalent Lagrangians in curved space
(involving for instance a minimal or a non-minimal coupling of matter to gravity, both cases admitting different invariances):
by considering both the diffeomorphism invariance and the absence or presence of \emph{local Weyl invariance}
in curved space, one can recover, in the flat space limit, the improved and the new improved EMT's for a scalar field discussed before.
Moreover, one can relate the corresponding symmetries in curved space and in flat space (local versus global scale
invariance)~\cite{Callan:1970ze, Deser:1970hs}.

Since the issues we just raised concern the relationship between results for field theories in Minkowski space-time (like~\eqref{eq:relatecurrents})
and results in general relativity (field theories in curved space-time), we mention that general relativity may also be viewed as a solution
to the problem of constructing a  field theory for a massless spin $2$ field (symmetric tensor field) interacting with matter in  Minkowski space-time,
 i.e. a consistent special relativistic field theory for a massless spin $2$ particle.
More precisely, starting from the free field theory for such a field in  Minkowski space-time as given by the Lagrangian of Pauli and Fierz (which amounts
to a linearization of the pure gravitational field equations), one has a local gauge invariance (corresponding to a linearization of diffeomorphisms)
and a consistent coupling to the EMT of matter then requires a back-reaction of the gravitational field on matter:
this process, referred to as \emph{Noether method},
has to be re-iterated in principle an infinite number of times.
However, by the use of a classically equivalent first-order action functional determined by J.~Schwinger,
S.~Deser could
promote the free field theory to a consistent self-interacting theory in a few steps~\cite{\deser}.
The initial flat space metric does not appear in the final theory which rather involves a symmetric tensor field $(g_{\m \n})$ representing
the dynamics of the gravitational field.
This non-geometrical approach to general relativity is described in detail in reference~\cite{Ortin} (see also~\cite{Feynman-book,\straumann})
and also played a role for YM-theories since R.~Feynman introduced the notion of ghost fields in this context before it has been considered by
B.~DeWitt and by
L.D.~Faddeev and V.N.~Popov for the quantization of YM-theories~\cite{YM50}.

\subsection{Fields in curved space-time}\label{sec:fieldsCurvedST}

\subsubsection{General framework and dynamics}

\paragraph{Curved space-time:}
Let $M$ be a \emph{$n$-dimensional space-time manifold}
endowed with a \emph{metric tensor field} $\mg  \equiv \left( g_{\mu \nu} \right)$
of signature\footnote{%
This signature is common in field theory
and is often used in general relativity too (e.g. in references~\cite{LandauLifschitz2,Birrell:1982ix,Ortin}),
though a large part
of the literature on general relativity uses the opposite signature $(-,+,\dots,+)$.
Different conventions for the signature
in combination with different sign conventions for the Riemann and Einstein tensors
imply differences of signs
in various mathematical expressions.
In particular, we choose here the ``$(-,+,-)$'' convention
for the curvature tensor
according to the classification of Misner, Thorne and Wheeler~\cite{Misner:1973}.} $(+,-,\dots,-)$.
We denote the \emph{covariant derivative} of a tensor field with respect to the Levi-Civita-connection
by $\nabla_{\m}$
(e.g.
$\nabla _{\m} V^{\rho} = \partial _{\mu} V^{\rho} + \Gamma^{\rho} _{\mu \n} V^{\n}$
where
the coefficients
$
\Gamma^{\rho} _{\m \n} \equiv \frac{1}{2} \, g^{\rho \sigma}
\left( \pa_\m g_{\n \sigma} +  \pa_\n g_{\m \sigma } -  \pa_\sigma g_{\m  \n} \right)
$
are the \emph{Christoffel symbols}) and we write
the \emph{gauge covariant derivative} as before by $D_{\m}$,
e.g. for a complex scalar field $D_{\mu} \phi \equiv
\nabla_{\mu} \phi + \ri \mq A_{\mu}  \phi = \pa_{\mu} \phi + \ri \mq A_{\mu}  \phi$.

The commutator of covariant derivatives defines the \emph{Riemann curvature tensor,} i.e.
\[
[\nabla_\m , \nabla_\n] V^\rho
= ?R^\rho_{\sigma \m \n}? V^\sigma
\,,
\]
and by a contraction of indices the latter gives rise to the \emph{Ricci tensor}
$R_{\m \n} \equiv ?R^\rho_{\mu \n \rho}?$,
which yields the \emph{curvature scalar} $R\equiv ?R^\m_\m?$.
The equations of motion for the gravitational field
involve the \emph{Einstein tensor} $G_{\m \n} \equiv R_{\m \n} - \frac{1}{2} \, g_{\m \n} R$
which is covariantly conserved $\nabla^\m G_{\m \n} =0$ as a consequence of its definition.
(Here and in the following, we limit ourselves for simplicity to the case
where the cosmological constant $\Lambda$ vanishes, otherwise an extra $\Lambda$-dependent term
appears in $G_{\m \n} $ and in the action functional for the gravitational field.)

\paragraph{Matter/gauge fields:}
The coupling of fermionic (spinorial)
matter fields to gravity requires the consideration of vielbein fields
and will be addressed in section~\ref{sec:spinfields}.
In the sequel we only consider bosonic (tensor)
fields, and more precisely scalar matter fields $\phi$ and/or
gauge vector fields $A\equiv (A^{\mu})$.
The YM field strength tensor in curved space is defined by
$$
F_{\m \n} \equiv  \nabla_{\m} A_{\n}-\nabla_{\n}A_{\m} + \ri \mq \, [ A_\m , A_\n ]
= \pa_{\m} A_{\n}-\pa_{\n}A_{\m} + \ri \mq \, [A_\m , A_\n ]
\, ,
$$
where the last expression follows from the symmetry of the Christoffel symbols
$(\Gamma_{\m \n} ^\rho = \Gamma_{\n \m} ^\rho )$.

The scalar matter fields $\phi$ and/or the gauge field $A$ will generically be denoted
 by $\vp$.  Their dynamics is assumed to be given by an
action functional $S_M [\vp , \mg] \equiv \int d^n x \, \sqrt{|g|} \, {\cal L}_M$
(involving $g \equiv \det \, \mg$)
which is invariant under diffeomorphisms,
as well as under gauge transformations $A_\m \leadsto A'_\m = A_\m + D_\m \omega$
if a gauge field is present (see equation~\eqref{eq:CDom}).
Thus, ${\cal L}_M$ is a scalar density with respect to diffeomorphisms.

\paragraph{Dynamics of fields:}
Following D.~Hilbert, the dynamics and interaction of all fields
is described by the total action
\begin{align}
\label{eq:totalactgrav}
S[\vp , \mg]  \equiv S_{\txt{grav}} [\mg] + S_M [\vp ,\mg]
\equiv \int d^nx  \, \sqrt{|g|} \, {\cal L}_{\txt{grav}} (R)
 + \int d^nx  \, \sqrt{|g|} \, {\cal L}_M ( \vp , \mg )
\, .
\end{align}
Here,  ${\cal L}_{\txt{grav}} (R) \equiv \frac1{2 \kappa} \, R$
(with $\kappa \equiv 8 \pi G$, where $G$ is Newton's constant),
and for the gauge field $A$ we have the Lagrangian density
\begin{align}
 {\cal L}_M (A, \mg )
 & \equiv - \frac{1}{4c_2}  \, \Tr \, (F^{\mu \nu} F_{\mu \nu})
 = -  \frac{1}{4} \, g^{\m \rho} g^{\n \sigma} F^a_{\rho \sigma} F^a_{\mu \nu}
 \, .
\label{eq:curvedspaceLag}
\end{align}
For a multiplet $\phi$ of complex scalar fields interacting with the gauge field,
the minimal coupling to gravity is described by the Lagrangian\footnote{More generally,
we can include a gauge invariant self-interaction potential $V(\phi^{\dagger} \phi)$:
this will not change the form~\eqref{eq:curvedspaceEMT} of the resulting EMT.}
\begin{align}
 {\cal L}_M  ( \phi , A, \mg )
  \equiv g^{\mu \nu} (D_\mu \phi^\dagger) ( D_\nu \phi ) - {m^2} \, \phi^\dagger \phi
 \, , \qquad \textrm{with} \ \;
  D_\mu \phi \equiv \pa_\m \phi + \ri \mq A_\m \phi
 \, .
\label{eq:curvedspaceMGLag}
\end{align}

Instead of the  \emph{minimal coupling} of the scalar matter field to gravity one may consider
\begin{align}
 \tilde{{\cal L}} _M (\phi , \mg ) \equiv  {\cal L}_M (\phi , \mg ) -
  \frac12 \, \xi R \phi^\dagger \phi
  \, ,
  \label{eq:LagMR}
 \end{align}
 where $\xi$ is a real parameter and where the second term represents a \emph{non-minimal coupling} of the matter field $\phi$
to gravity: in fact this is the only possible local scalar coupling with the correct
dimension~\cite{Birrell:1982ix, Wald:1984}.

The action~\eqref{eq:totalactgrav}, which defines the dynamics of fields in \emph{general relativity,}
is invariant under general coordinate transformations.
The  equations of motion for the interacting matter and gauge fields following from the functionals~\eqref{eq:totalactgrav}-\eqref{eq:curvedspaceMGLag}
read $0= \delta S[\vp , \mg] /\delta \vp = \delta S_M [\vp , \mg] /\delta \vp$ which yields
\begin{align}
0=& \,  D_\m F^{\m \n} - j^\n
\qquad \qquad \quad \ \mbox{(YM equation in curved space),}
\nn
\\
0=& \, (D^\m D_\m +m^2) \phi =0
\qquad \mbox{(Coupled Klein-Gordon equation in curved space).}
 \end{align}
Here, $D_\m F^{\m \n} = \nabla_\m F^{\m \n} + \ri \mq [A_\m, F^{\m \n} ]$ where
$\nabla_\m F^{\m \n} = \frac{1}{\sqrt{|g |}} \, \pa_\m \left( \sqrt{|g |}
\, F^{\m \n} \right)$ denotes the covariant divergence of the antisymmetric tensor $F^{\m \n}$.
The derivative $D^\m D_\m \phi$ involves the contribution
$\Box \phi \equiv g^{\m \n} \nabla_\m \nabla_\nu \phi =
\frac{1}{\sqrt{|g |}} \, \pa_\m \left( \sqrt{|g |}
\, g^{\m \n} \pa_\n \phi \right)$,
i.e. the \emph{Laplace-Beltrami operator} acting on scalar fields.
The equations of motion of the  metric field components represent \emph{Einstein's field equations:}
\begin{align}
0 = \frac{\delta S}{\delta g^{\mu \nu}} =  \frac{\delta S_{\txt{grav}}}{ \delta g^{\mu \nu}} + \frac{\delta S_M }{ \delta g^{\mu \nu}}
= \frac{\sqrt{|g|} }{ 2 \kappa} \, \left( G_{\mu \nu} + \kappa T_{\mu \nu} \right)
\, ,
\qquad \txt{i.e.} \ \;
\Boxed{
G_{\mu \nu} = - \kappa T_{\mu \nu}
}
\label{eq:Einsteineq}
\end{align}
with
\begin{align}
 \Boxed{
T^{\mu \nu}  [\vp, \mg] \equiv
\frac{-2}{\sqrt{|g|}}
\,  \frac{\delta S_M [\vp , \mg ] }{\delta g_{\mu \nu} }
}
\, .
 \label{eq:metricEMT}
\end{align}
For obvious reasons, we will refer to this tensor
as the \textbf{metric EMT} (in curved space).
By reference
to its originators~\cite{Hilbert:1915tx, \einstein} it is also called the \emph{Einstein{-}Hilbert EMT in curved space.}
Concerning the global sign in~\eqref{eq:metricEMT} we note that $g^{\m \n} g_{\n \lambda} = \delta^\m _{\lambda}$ implies that
$T_{\mu \nu}  =
\frac{+2}{\sqrt{|g|}} \,  \frac{\delta S_M [\vp , \mg ]}{\delta g^{\mu \nu} }$.

\subsubsection{\EMT for matter/gauge fields coupled to a dynamical gravitational field}\label{sec:dynmetric}

\paragraph{On the metric EMT:}
From the equations of motion~\eqref{eq:Einsteineq} one concludes that the local field $ T^{\mu \nu}  [\vp, \mg] $ plays a fundamental role
if matter or gauge fields are coupled to gravity
while the formulation of field theory in flat space
essentially relies on the conserved charges $P^\n \equiv \int d^{n-1}x \, T_{\txt{can}}^{0 \n}$.
Since the metric tensor is  symmetric in its indices, the metric \EMT is
\emph{identically symmetric by construction,}
i.e. we have a symmetric expression without using the equations of motion of $\vp$.
From $\nabla_\m G^{\m \n} =0$ it follows that the EMT is \emph{covariantly conserved} in curved space,
i.e.
\begin{align}
\label{eq:CovConsLawEMT}
\Boxed{
\nabla_\m T^{\m \n} =0
}
\,,
\end{align}
if $\mg$ satisfies Einstein's field equations.
Due to the presence of the covariant rather than the ordinary derivative in equation~\eqref{eq:CovConsLawEMT},
this relation does not represent a local conservation law.
This can be understood on physical grounds due to the fictitious forces which appear
in arbitrary (accelerated) frames~\cite{\einstein, Thirring:1997}.
However, relation~\eqref{eq:CovConsLawEMT} can be related
to  a local conservation law and, in certain instances, integral conservation laws for energy and momentum
can be derived, e.g. see
references~\cite{LandauLifschitz2, Thirring:1997, Abbott:1981ff, DuboisViolette:1986ws, Barnich:2001jy, StefaniGR, Ortin}.

\paragraph{Explicit expressions:}
From $S_M  [\vp , \mg ] \equiv \int d^nx \, \sqrt{|g|} \; {\cal L}_M (\vp, \pa_\mu \vp; \mg)$ and the definition~\eqref{eq:metricEMT}
we can obtain an explicit expression of $T^{\mu \nu}$ in terms of the Lagrangian density ${\cal L}_M$ (the latter depending on the first order derivatives
$\pa_\mu \vp$):
\begin{align}
\Boxed{
T^{\mu \nu}  = -2 \,
\frac{\pa {\cal L}_M}{\pa g_{\mu \nu}} - g^{\mu \nu} {\cal L}_M
}
\label{eq:genEMTLag}
\,.
\end{align}
For our prototype Lagrangians~\eqref{eq:curvedspaceLag}
and~\eqref{eq:curvedspaceMGLag}
we get the results
\begin{align}
T^{\m \n} [A, \mg]
&=  \frac{1}{c_2} \, \Tr \big( F^{\m \rho} {F_{\rho}}^{\nu}
+ \frac14 \, g^{\mu \nu} F^{\rho \sigma} F_{\rho \sigma} \big)
 \, ,
\label{eq:curvedspaceEMT}
\\
 T^{\m \n} [\phi , A, \mg]
&=(D^\mu \phi^\dagger ) ( D^\nu \phi ) +
( D^\nu \phi ^\dagger ) (D^\mu \phi )
-     g^{\mu \nu} {\cal L}_M ( \phi, A, \mg )
\, .
 \nonumber
\end{align}
For the Lagrangian~\eqref{eq:LagMR}
the tensor $ T^{\m \n} [\phi , \mg]$ involves additional contributions,
see references~\cite{Birrell:1982ix, Wald:1984}.
The conservation law $\nabla_\m T^{\m \n} =0$ can be explicitly checked
for expressions~\eqref{eq:curvedspaceEMT}
by using the equations of motion
of $A$ and $\phi$ as well as the Bianchi identity $0 = D_{\rho} F_{\m \n} + $
cyclic permutations of the indices $\rho, \m , \n$.
The metric EMT $T^{\mu \nu}  [A, \mg] $
is gauge invariant as a consequence of the gauge invariance of the Lagrangian
density ${\cal L} _M ( A, \mg ) $
and of definition~\eqref{eq:genEMTLag}. The results concerning the tracelessness of the EMT's~\eqref{eq:curvedspaceEMT}
are the same as the ones obtained for the improved EMT's in Minkowski space.

\paragraph{Invariance under diffeomorphisms and the metric EMT:}
Let us verify that  \emph{the covariant conservation law $\nabla_\m T^{\m \n} =0$ follows from the diffeomorphism invariance of the total
action~\eqref{eq:totalactgrav}}. To do so, we note that
a diffeomorphism $x^\m \leadsto x'^\m (x) \simeq x^\m - \xi^\m (x)$ is generated by a smooth vector field $\xi \equiv \xi^\m \pa_\m$ and acts on the
metric tensor field as
$\delta_{\xi} g_{\m \n} = \nabla_\m \xi_\n + \nabla_\n \xi_\m $.
From the invariance of the action under infinitesimal diffeomorphisms and the use of the equations of motion of matter and gauge fields
it follows that
\begin{align}
0 = \delta_{\xi} S = \int d^nx \, \left( \frac{\delta S}{ \delta \vp} \, \delta_{\xi} \vp +  \frac{\delta S}{ \delta  g_{\m \n}} \, \delta_{\xi}  g_{\m \n} \right)
=
\int d^nx \, \frac{\delta S}{ \delta  g_{\m \n}} \, 2 \nabla_\m \xi_\n
\, ,
 \label{eq:ConseqDiffInv1}
\end{align}
where we took into account the symmetry of $g_{\m \n}$ for passing to the last expression.
By substituting the explicit form of the functional derivative $\frac{\delta S}{ \delta  g_{\m \n}}$
as given in equation~\eqref{eq:Einsteineq} and performing an integration by parts, we obtain
 \begin{align}
0 = \delta_{\xi} S = - \frac{1}{\kappa} \int d^nx \, \sqrt{|g|} \, ( G^{\m \n} + \kappa T^{\m \n} ) \nabla_\m \xi_\n
= \frac{1}{\kappa} \int d^nx \, \sqrt{|g|} \; \xi_\n    ( \nabla_\m  G^{\m \n} + \kappa  \nabla_\m T^{\m \n} )
 \label{eq:ConseqDiffInv2}
\, .
\end{align}

First, we consider the particular case where we have pure gravity, i.e. no matter fields, hence $T^{\m \n} =0$.
From the arbitrariness of $\xi_\n$ one then concludes that $  \nabla_{\m}G^{\mu \nu} =0$.
In fact~\cite{\straumann},
this line of arguments may be viewed as indirect proof of the relation  $  \nabla_{\m}G^{\mu \nu} =0$
which holds by virtue of the definition of $G^{\mu \nu}$ in terms of the metric.
The latter relation is referred to as generalized Bianchi identity or as \emph{Noether identity} since it follows from the invariance
of the action functional under a group of \emph{local} symmetry transformations,
i.e. an illustration of \emph{Noether's second theorem}~\cite{Noether:1918zz}.

Next we consider the total action: by using the identity  $  \nabla_{\m}G^{\mu \nu} =0$ and the arbitrariness of $\xi_\n$,
it then follows that the relation $  \nabla_{\m}T^{\mu \nu} =0$ holds for the solutions of the matter field equations.

\subsubsection{\EMT for matter/gauge fields coupled to an external gravitational field}\label{sec:extmetric}

\paragraph{Generalities:}
If we consider matter and gauge fields coupled to an external (background) metric, then we do not have a dynamical
term for gravity in the action, i.e. $S = S_M [\vp; \mg ]$. The EMT now represents the variation of the total action with respect to the
external gravitational field:
\begin{align}
\Boxed{
T^{\mu \nu}  [\vp; \mg] \equiv
\frac{-2}{\sqrt{|g|}}
\,  \frac{\delta S [\vp ; \mg ] }{\delta g_{\mu \nu} }
}
\qquad \mbox{($\mg =$ external gravitational field)}
 \, .
 \label{eq:extmetricEMT}
\end{align}
The covariant conservation law  $  \nabla_{\m}T^{\mu \nu} =0$ for the solutions of the matter field equations
again follows from the calculation~\eqref{eq:ConseqDiffInv1}, \eqref{eq:ConseqDiffInv2} in which we now drop the contribution
$ S_{\txt{grav}} [ \mg ]$ to the action, which implies that $G^{\m \n}$ does not appear in the integral~\eqref{eq:ConseqDiffInv2}.

\paragraph{Case of an external gauge field:}

Let us also consider the case where both the metric field $\mg$ and the gauge potential  $(A^\m)$
represent external fields. We suppose that the complex scalar field $\phi$ is coupled
to the background gauge field $A$, i.e. we have the gauge invariant total action $S \equiv S_M [ \phi; A, \mg]
\equiv \int d^n x \, {\sqrt{|g|}} \, {\cal L}_M$ with ${\cal L}_M$ given by expression~\eqref{eq:curvedspaceMGLag}.
Then the gauge invariance of the total action $S$ and the use of the scalar
matter field equations of motion
$\delta S/\delta \phi =0$
imply the covariant conservation law $D _\m j^\m =0$ for the  current density vector
$(j^\m _a)$ associated to the scalar field; indeed, for an infinitesimal gauge variation
($\delta_{{g}} A_\mu =  D_\m \omega$)
we have
\[
0 = \delta_{{g}} S = \int d^nx \,
 \left( \frac{\delta S}{\delta \phi} \, \delta_{{g}} \phi
+ \frac{\delta S}{\delta A_{\mu}} \, \delta_ {{g}} A_{\mu} \right)
= \int  d^nx \, \frac{\delta S}{\delta A_{\mu}} \, D_\m \omega
=
 \int  d^nx \, {\sqrt{|g|}} \, \omega^a ( D_\m j^\m)_a
\, ,
\]
where
\[
 \Boxed{
j^\m _a \equiv
\frac{-1}{\sqrt{|g|}}
\, \frac{\delta S}{\delta A_{\mu}^a}
}
\qquad \mbox{($A, \mg =$ external fields)}
\, .
\]
For the matter field Lagrangian~\eqref{eq:curvedspaceMGLag} we obtain the explicit expression
\begin{align}
j^\m _a [\phi; A, \mg ] = \ri \mq \Big[  \phi^\dagger {T}_a D^\mu \phi
-  (D^{\mu } \phi^\dagger) {T}_a  \phi \Big]
\, .
\label{eq:CurDensCurv}
\end{align}

The invariance of the total action $S$ under diffeomorphisms then leads to the \emph{continuum version
of the Lorentz-Yang-Mills force law} in curved space-time~\cite{Balasin:2014dma},
\begin{align}
 \Boxed{
 \nabla_\m T^{\m \n} [\phi ; A, \mg ] = \frac{1}{c_2} \, \Tr \, (F^{\n \m} j_\m )
}
\label{eq:ContLorforce}
\end{align}
by virtue of
\[
0 = \delta_{\xi} S
=  \int d^nx \, \left( \frac{\delta S}{\delta \phi} \, \delta_{\xi} \phi
+  \frac{\delta S}{\delta A_\m} \, \delta_{\xi} A_\m
+ \frac{\delta S}{\delta g_{\mu \nu}} \, \delta_{\xi} g_{\mu \nu}
\right)
\, ,
\]
and $\delta S/\delta \phi =0, \, \delta_{\xi} g_{\m \n} = \nabla_{\m}\xi_{\n}+\nabla_{\n}\xi_{\m}$
as well as
\[
\delta_{\xi} A_\m =\xi^\n \pa_\n A_\m + (\pa_\m \xi^\n) A_\n = \xi^\n F_{\n \m} + D_\m (\xi^\n A_\n )
\, .
\]
For instance, for the matter field Lagrangian~\eqref{eq:curvedspaceMGLag}, we obtain the
explicit expression~\eqref{eq:curvedspaceEMT} for $T^{\m \n} [\phi ; A, \mg ]$ and~\eqref{eq:CurDensCurv}
for $ j^{\m} [\phi ; A, \mg] $.
Once the external gauge field is promoted to a
\emph{dynamical gauge field}
by adding the gauge field Lagrangian ${\cal L}_M (A; \mg )$ of equation~\eqref{eq:curvedspaceLag}
to the interacting matter field Lagrangian~\eqref{eq:curvedspaceMGLag}, the total  EMT of both matter and gauge fields is covariantly conserved
as we noted already after equation~\eqref{eq:extmetricEMT}.
In the flat space limit, expression~\eqref{eq:curvedspaceEMT} for $T^{\m \n} [\phi ; A, \mg ]$
reduces to expression~\eqref{eq:IntscalEIT} for the EMT  $T_{\txt{int}}^{\mu \nu} [ \phi ,A ]$
and equation~\eqref{eq:ContLorforce} reduces to the balance equation~\eqref{eq:divEMTs}
for the latter EMT.

\subsection{Einstein{-}Hilbert's \EMT for matter/gauge fields in Minkowski space}\label{sec:EHtensor}

\paragraph{Definition:}
The result \eqref{eq:extmetricEMT}  serves as a motivation for
defining the \emph{Einstein{-}Hilbert}
or \emph{metric EMT
for bosonic
matter fields $\vp$ in flat space} by the relation
\begin{align}
 \Boxed{
T_{\txt{EH}}^{\mu \nu} [\vp ] \equiv
\bigg(
\frac{-2}{\sqrt{|g|}}
\,  \frac{\delta S[\vp ; \mg ] }{\delta g_{\mu \nu}} \bigg) \! \bigg|_{\mg = \eta}
}
\, ,
 \quad \txt{where} \ \;
 \left\{
 \begin{array}{l}
\mg =  \mbox{external gravitational field}
\\
\eta \equiv (  \eta_{\mu \nu}) \equiv \txt{diag} \, (1, -1,\dots ,-1 ) \, .
 \end{array}
 \right.
\label{eq:HEMT}
\end{align}
By construction,
the so-defined \EMT is \emph{identically symmetric} and it
\emph{admits a natural generalization to curved space} by its very definition.
The flat space
\emph{conservation law} $\pa_\m T_{\txt{EH}}^{\mu \nu} =0$ for $T_{\txt{EH}}^{\mu \nu}$
follows directly from the
covariant conservation law $\nabla_\m T^{\mu \nu} =0$ in curved space.

The definition~\eqref{eq:HEMT} for the EMT of a physical system in Minkowski space
amounts to coupling this system to a gravitational field: then
$T_{\txt{EH}}^{\mu \nu} [\vp ]$ \emph{represents the response of the system to a variation
of the (external) metric to which all (energy and momentum carrying)  fields couple and which mediates the gravitational interaction
of these fields.}
This definition of the EMT in Minkowski space is conceptually and mathematically quite different from the one of $T_{\txt{imp}}^{\mu \nu} [\vp ]$
which we presented  in section~\ref{sec:EMTlag} and which follows from Noether's theorem (eventually supplemented by an improvement procedure
to render the canonical expression of the EMT symmetric in its indices
or gauge invariant, or both symmetric and traceless).
Before trying to relate the different expressions for the EMT's, we have a look at the explicit expressions for the
Einstein-Hilbert EMT.

\paragraph{Explicit expressions:}
The YM Lagrangian ${\cal L}_M (A, \mg )$ in curved space, as given by~\eqref{eq:curvedspaceLag},
 reduces in the flat space limit to the YM Lagrangian~\eqref{eq:invact}
 in Minkowski space;
similarly both of the scalar field Lagrangians ${\cal L}_M (\phi, \mg )$ and $\tilde{{\cal L}} _M (\phi, \mg )$
given in equations~\eqref{eq:curvedspaceMGLag} and~\eqref{eq:LagMR} reduce in the flat space limit
to one and the same Minkowski space Lagrangian ${\cal L}( \phi )$.
Furthermore,  the metric EMT's~\eqref{eq:curvedspaceEMT}
for the YM field and for the scalar field coupled to the former field reduce in the flat space limit to
the improved EMT's~\eqref{eq:impEMTYM} and~\eqref{eq:IntscalEIT}.
We also mention that for the non-minimal coupling of a free massless scalar field to gravity
as described by the Lagrangian~\eqref{eq:LagMR}, the corresponding metric EMT reduces in the flat
 space limit to the new improved EMT~\eqref{eq:confEMT}.

\paragraph{Relating the different expressions:}

In the following, we will show that the two
definitions for the EMT's of YM-theories in Minkowski space, i.e.~\eqref{eq:HEMT} which results from the coupling to gravity,
and the improved EMT~\eqref{eq:TotintEIT}
(with the definitions~\eqref{eq:IntEIT} and~\eqref{eq:IntEITmatter})
which follows from Noether's first theorem supplemented by the ``gauge improvement'' procedure,
coincide  with each other.
(We refer to the work~\cite{Forger:2003ut}
for general mathematical arguments and results relying on differential geometric tools.)
The coincidence of results for the EMT's can readily be explained on general grounds
by comparing the expression~\eqref{eq:genEMTLag} for the metric EMT,
i.e. $ T^{\mu \nu}  = -2 \,
\frac{\pa {\cal L}_M}{\pa g_{\mu \nu}} - g^{\mu \nu} {\cal L}_M$,
with the general expressions~\eqref{eq:IntEIT}, \eqref{eq:IntEITmatter}
for the improved EMT's of the YM field $A$ and of a scalar field multiplet $\phi$
coupled to the YM field.

 First, we note that the second term in $T^{\mu \nu} $ reduces directly  to the second term in $T_{\txt{int}}^{\mu \nu}$
as $g^{\m \n}$ reduces to $\eta^{\m \n}$.
Concerning the first term in $T ^{\mu \nu}$, i.e.  $-2 \, \frac{\pa {\cal L}_M}{\pa g_{\mu \nu}}$,
 we note that the curved space
 Lagrangians  ${\cal L}_M (A, \mg )$ and ${\cal L}_M (\phi, A, \mg )$
 (as given by expressions~\eqref{eq:curvedspaceLag}, \eqref{eq:curvedspaceMGLag})
 are quadratic in the
field strengths ($F_{\m \n}$ and $D_\m \phi$, respectively)
with coefficients depending on $ g^{\rho \sigma}$. Thus the relation
\begin{align}
\frac{\pa g^{\rho \sigma}}{\pa g_{\m \n}} = - g^{\rho \alpha} g^{\sigma \beta} \delta^{(\m \n )}_{(\alpha \beta)}
\, ,
\qquad \txt{with} \ \;
\delta^{(\m \n )}_{(\alpha \beta)} = \frac{\pa g_{\rho \sigma}}{\pa g_{\m \n}}
\, ,
\label{eq:DeriveG}
\end{align}
implies that $-2 \, \frac{\pa {\cal L}_M}{\pa g_{\mu \nu}}$ reduces in the flat space limit to
an expression which is again quadratic  in the field strengths and which is, by construction,
Lorentz covariant, gauge invariant and symmetric in the indices. By virtue of Euler's homogeneous function theorem
we have
\begin{align}
-2 \, \frac{\pa {\cal L}_M  }{\pa g_{\mu \nu}} (A, \mg ) \ \ \stackrel{\mg \to \eta}{\longrightarrow} \ \ &
2 \, \frac{\pa {\cal L}_g}{\pa F^a _{\m \rho}} \, {F_a ^\n}_\rho
\,,\\
-2 \, \frac{\pa {\cal L}_M  }{\pa g_{\mu \nu}} (\phi , A, \mg )\ \ \stackrel{\mg \to \eta}{\longrightarrow} \ \ &
\frac{\pa {\cal L}_M}{\pa (D_{\mu} \phi )} \, D^\n \phi
+ ( D^\n \phi^\dagger ) \, \frac{\pa {\cal L}_M}{\pa (D_{\mu} \phi^\dagger )}
\, ,
\nonumber
\end{align}
i.e. the results~\eqref{eq:IntEIT} and~\eqref{eq:IntEITmatter}.

\subsection{Case of spinor fields}\label{sec:spinfields}
The coupling of spinor fields to gravity
(put forward in Weyl's seminal  paper~\cite{ORaifear} of 1929)
requires the consideration of orthonormal
vielbein fields ${e^a}_\m (x)$
related to the metric by $g_{\m \n} = \eta_{ab} {e^a}_\m {e^b}_\n$.
The components ${E^\m}_a$
of the inverse of the matrix $({e^a}_\m)$ define the frame vector field $E_a \equiv
{E^\m}_a \pa_\m$: ${E^\m}_a
{e^a}_\n = \delta_\m ^\n$. We have $\sqrt{|g|} =e$ with  $e \equiv |\det \, \left( {e^a}_\m \right) \! |$.
The
\emph{vielbein EMT}~\cite{Ortin} is then defined by varying the action
for the spinor fields
(coupled to an \emph{external} gravitational field)
with respect to the vielbein or frame fields:
\begin{align}
 \Boxed{
{T^a}_\m = \frac{1}{e} \,
\frac{\delta S}{\delta {E^\m}_a }
}
\, , \qquad \txt{or} \ \
{T^\m}_a = \frac{-1}{e} \, \frac{\delta S}{\delta {e^a}_\m } \,
\label{eq:spinEMT}
\,.
\end{align}
The invariance of the action $S [\vp,  {e^a}_\m ]$ under \emph{local Lorentz transformations}
(parametrized at the infinitesimal level by $\delta_{\varep}  {e^a}_\m = {\varep ^a}_b  {e^b}_\m $
with $\varep_{ab} = - \varep_{ba}$)
and the application of the matter field equations of motion $\delta S / \delta \vp =0$
 imply that the tensor $T^{ab}$ is \emph{symmetric on-shell:} from
\begin{align}
0 = \delta_{\varep} S = \int d^nx \, \left( \frac{\delta S}{ \delta \vp} \, \delta_{\varep} \vp
+  \frac{\delta S}{ \delta {e^a}_\m } \, \delta_{\varep} {e^a}_\m  \right)
=
- \int d^nx \, e \, {T^\m}_a \, {\varep ^a}_b  {e^b}_\m
= \frac{1}{2} \int d^nx \, e \, T^{[ab]} \varep_{ab}
\, ,
 \label{eq:LorInv}
\end{align}
and from the arbitrariness of $\varep_{ab}$ we conclude that $T^{[ab]} =0$, or equivalently $T^{ab} = T^{ba}$.
Then, the spinor field EMT  $T^{\m \n} \equiv {E^\m}_a {E^\n}_b T^{ab}$ is also symmetric
in the curved space indices
for the solutions of the matter field equations.
Furthermore~\cite{Bertlmann-book}, the matter field equations ensure the covariant conservation law $\nabla_\m T^{\m \n} =0$
by virtue of the line of arguments~\eqref{eq:ConseqDiffInv1}, \eqref{eq:ConseqDiffInv2} with $S \equiv S_M$.

In summary, the matter field equations for the spinor fields ensure the consistency of Einstein's field equations $G^{\m \n} = - \kappa T^{\m \n}$ involving
the Einstein tensor $G^{\m \n}$ which is both symmetric and covariantly conserved
in the absence of torsion~\cite{Ortin}. (The latter assumption is generally made in Einstein gravity, but we note that
a generalization of the theory including torsion is given by the so-called \emph{Cartan-Sciama-Kibble  approach to gravity,}
see~\cite{\hehlhammond, Ortin} and references therein.)
Accordingly, the EMT with lower indices (which also has to be on-shell symmetric in the absence of torsion) is generally written as
\begin{align}
 \Boxed{
T_{\m \n} = \frac{1}{2} \, ( {e^a}_\m \eta_{ab} {T^b}_\n + ( \mu \leftrightarrow \nu )  )
}
\label{eq:symspinEMT}
\,.
\end{align}

By way of illustration, we
consider the  \textbf{Dirac field} coupled to a gauge field $(A^\m )$ and to gravity.
Then we have covariant derivatives involving the  spin connection  $\omega_\mu$,
\begin{align}
\label{eq:CovDerApsi}
D_a \psi \equiv {E^\m}_a D_\m \psi
=  {E^\m}_a  [ \pa_\m \psi + \omega_\m  \psi  + \ri \mq A_\m \psi ]
\,,\qquad
D_a \bar\psi={E^\m}_a  [ \pa_\m \bar\psi - \bar\psi \,\omega_\m  - \ri \mq \bar{\psi} A_\m ]
\,,
\end{align}
and the action functional reads\footnote{We note that the inclusion of an invariant
Yukawa-like coupling with scalar fields does not modify the form of the resulting EMT.}
\begin{align}
S[ \psi , A ; {e^a}_\m ] & \equiv \int d^nx \, e \, \bar{\psi} \left( \ri
\gamma^a \stackrel{\leftrightarrow}{D}_a \psi - m \psi \right)
\, , \quad
\nonumber
\\
& =  \int d^nx \, e \, \Big[ \frac{\ri}{2} \, \big( \bar{\psi}
\gamma^a D_a \psi
- (D_a \bar{\psi}) \gamma^a  \psi \big)
- m \bar{\psi} \psi \Big]
\label{eq:actDi}
\\
& \equiv  \int d^nx \, e \, {\cal L}_M  (\psi , A , {E^\m}_a  )
\nn
\, .
\end{align}
Variation of this action with respect to the frame fields
${E^\m}_a  $
and use of the matter field equations of motion (which imply that the term
in $\delta S$ which is proportional to $\delta e$,
i.e. the function ${\cal L}_M$, vanishes) yields
\begin{align}
 \Boxed{
 {T^a}_\m =\frac{\pa {\cal L}_M}{\pa {E^\m}_a }
}
\label{eq:EMTDirL}
\, ,
\end{align}
hence ${T^a}_\m  = \ri \bar{\psi} \gamma^a \stackrel{\leftrightarrow}{D}_\m \psi$.
The associated symmetric tensor \eqref{eq:symspinEMT} now has the form
\begin{align}
 \Boxed{
T^{\m \n} =
\frac{\ri}{2} \, \big(  \bar{\psi} \gamma^\m \stackrel{\leftrightarrow}{D}\!^\n \psi
+ \bar{\psi} \gamma^\n \stackrel{\leftrightarrow}{D}\!^\m \psi \big)
}
\label{eq:symDirEMT}
\, .
\end{align}
By virtue of the matter field equations,
it is covariantly conserved and
coincides in the flat space limit with the \emph{improved \EMT~\eqref{eq:IntDiracEMT}
for the Dirac field coupled to the YM field}
(the latter being conserved, symmetric and gauge invariant).

\subsection{Summary}\label{sec:SummaryGrav}

In this section we recalled the definition of the metric EMT in curved space
and, for gauge field theories, we showed  that the resulting expressions reduce in the flat space limit
 to gauge invariant, symmetric EMT's which coincide with the improved EMT's determined in section~\ref{sec:EMTlag}.
 (For the case of a non minimal coupling of scalar fields to gravity, one recovers the new improved
 EMT in Minkowski space, the local scale invariance reducing to  global scale invariance.)

\section{On the quantum theory: Noether's theorem and Ward identities}

In the perturbative approach to Lagrangian models in quantum field theory,
the first Noether theorem, as applied to geometric symmetries (e.g. translational invariance)
or to internal symmetries (e.g. global $U(1)$ transformations), finds its expression
in the so-called Ward (or Ward-Takahashi)  identities.
A general formulation and simple derivation of the latter has been put forward
by R.~Stora~\cite{Rouet:1972ut, Rouet:1972bis, Stora:1973dka}
towards 1971. Here, we outline the general ideas which are nicely summarized in~\cite{Zuber:2013rha}
and refer to the monographs~\cite{Itzykson:2005, Srednicki:2007qs, DiFrancesco:1997nk}
for the technical details and physical applications.

We consider the case of a global internal symmetry transformation of a matter multiplet
$\vp \equiv [\vp_1, \dots , \vp_N ]^t$,
e.g. relations~\eqref{eq:globgaugetr}
with constant infinitesimal symmetry parameters
$\omega^a$ (with $a=1, \dots , n_G$);
thus, we have $\delta \vp^r (x) = \omega^a \Phi^r_a ( \vp (x))$  and \emph{Noether's
first theorem} then reads
\[
 \Phi^r_a \, \frac{\delta S}{\delta \vp^r} + \pa_\m j^\m _a =0 \, , \qquad \txt{with} \ \
 j^\m _a \equiv \frac{\pa {\cal L}}{\pa (\pa_\m \vp^r)} \,  \Phi^r_a
 \, .
\]
In quantum theory one is interested in the \emph{vacuum expectation values}
$\langle T \vp_{s_1} (y_1) \dots \vp_{s_n} (y_n)   \rangle$
of time-ordered field operators  $\vp_{s_1} (y_1), \dots , \vp_{s_n} (y_n) $
with $n=1, 2, \dots$.
Use of the  conservation law $\pa_\m j^\m _a =0 $ (associated to Noether's
theorem in the classical theory) and of the definition of the $T$-product in terms
of Heaviside's function lead to
\[
\frac{\pa \; }{\pa x^\m} \langle T  j^\m _a (x) \vp_{s_1} (y_1) \dots \vp_{s_n} (y_n)   \rangle
= \sum_{j=1}^n \delta (x^0 - y^0_j) \, \langle T \vp_{s_1} (y_1) \dots
[j^0_a (x) , \vp_{s_j} (y_j)] \dots
\vp_{s_n} (y_n)   \rangle
\, .
\]
By virtue of the canonical commutation relations for the field operators,
one can evaluate the equal-time commutator on the right-hand side: $\left. [j^0_a (x) , \vp_{s_j} (y_j)] \right|_{x^0 = y^0_j}
= - \ri \Phi^{s_j}_a (x) \, \delta(\vec x - \vec y_j )$. This yields
 the \emph{Ward identities} (in configuration space):
\begin{align}
\Boxed{
\frac{\pa \ }{\pa x^\m} \langle T  j^\m _a (x) \vp_{s_1} (y_1) \dots \vp_{s_n} (y_n)   \rangle
= - \ri \sum_{j=1}^n \delta (x - y_j) \, \langle T \vp_{s_1} (y_1) \dots \Phi^{s_j}_a (x)
 \dots \vp_{s_n} (y_n)   \rangle
}
 \, .
 \label{eq:WI}
\end{align}
By Fourier transformation, these identities can be rewritten in momentum space
and in fact it is in the latter space that they were originally discovered for a special case
in electrodynamics, and that they are often encountered in the literature.
Their general form~\eqref{eq:WI} shows that \emph{the Ward identities are the reflection of Noether's first
theorem: they represent a collection of
important relations between the correlation functions  in quantum field theory which result from the underlying
classical symmetry group.}

One may wonder about the impact of an improvement $j_a^\m \leadsto j_a^\m + \pa_\rho
B_a^{\rho \m}$ for a given locally conserved classical current density $(j_a^\m)$ on the explicit
expression of Ward identities.
General statements require a specification of the class of superpotentials which is considered.
If the latter depend on the canonical momenta, derivatives of the delta function potentially appear
in the Ward identities.
Here, we only refer to some general recent works devoted to Ward identities~\cite{Duetsch:2001sw, Avery:2015rga}.
Furthermore, for completeness, we mention
some works which deal more specifically with the quantum theory
related to the EMT (some others being discussed in the next
section):~\cite{Coleman-Lectures, Fulling:1989nb, Duff:1993wm, DiFrancesco:1997nk,Blumenhagen:2013fgp, \dymarskynakayama, Blaschke:2014ioa}.

\section{Ward identities, their ``cousins and descendants'': on (the work of) Raymond Stora}
The so-called Slavnov (or Slavnov-Taylor)
 identities can be viewed as a generalization of the Ward identities
and can be formulated and derived in simple terms using the so-called \emph{BRST symmetry}.
This symmetry has been discovered by  Becchi, Rouet and Stora (BRS)~\cite{\becchi, Becchi:1975nq}
(and shortly thereafter in an unpublished work by Tyutin~\cite{Tyutin:1975qk}) and it was shown
by BRS~\cite{\becchi, Becchi:1975nq} that this symmetry allows to prove the renormalizability of non-Abelian gauge theories
and to characterize the observables as well as the anomalous breakings of classical symmetries in quantum theory
(see references~\cite{Becchi:1975dw, Stora:1976kd, Becchi:1981jx, Piguet:1980nr, Becchi:1985bd} for some early elaborations
and the monograph~\cite{Piguet:1995} for an introductory account).
The BRST symmetry transformations were originally expressed in terms of an anti-commuting (Grassmann) parameter, but
BRS could quickly dispense with this parameter~\cite{Becchi:1975nq},
 thus regarding the symmetry operator as an anti-derivation
acting on the \emph{BRS differential algebra.}
R.~Stora always denoted the BRST-transformation of a field $\vp$ by $s \vp$ while referring to it as the
\emph{``Slavnov operation''}~\cite{Stora:1976kd,\stora}.
Due to its nilpotency, this operation allows for a cohomological interpretation~\cite{Becchi:1975nq, Stora:1976kd}
and thereby allows to reformulate the problem of perturbative renormalization of Lagrangian field theories
with rigid or local symmetries as an algebraic problem: this approach is referred to as \emph{algebraic renormalization},
see~\cite{FrancoPiguet} for a nice summary and references \cite{Piguet:1995,Schweda-book:1998} for a detailed presentation
and various applications.
In particular, the realization of Wess and Zumino~\cite{Wess:1971yu} that \emph{anomalies}
(which describe an anomalous breaking of classical symmetries in quantum theory)
are described by a \emph{consistency condition} could be neatly reformulated as a cohomological problem~\cite{Stora:1976kd}
(and treated by BRS for YM theories~\cite{Becchi:1975nq}).
In fact, the issue of determining the anomalous terms can be dealt with by using
a simple algebraic method (referred to as the \emph{descent equation method})
put forward by R.~Stora~\cite{Stora:1976kd} with the help of some mathematical lemmas due to J.A.~Dixon
(see also ~\cite{Zumino:1983ew, Stora:1983ct, Manes:1985df} as well as the general introduction
presented in the monograph~\cite{Bertlmann-book}).
This approach has been applied by
R.~Stora and his collaborators to a wealth of theories like gravity~\cite{Langouche:1984gn},
supersymmetry~\cite{\GirardiFerrara},
string theory~\cite{Baulieu:1986hw},
conformal models~\cite{\lazzarini}
or topological theories~\cite{Ouvry:1988mm},
and has been investigated by numerous authors.
Quite generally, the method of \emph{BRST quantization} within the Lagrangian or the Hamiltonian~\cite{Henneaux:1992} framework,
and its variants developed by the Russian School (Batalin, Fradkin, Tyutin, Vilkovisky),
represents a general and powerful approach to the quantization of constrained dynamical systems with a finite or infinite number
of degrees of freedom~\cite{Henneaux:1992, Gitman:1990,  Rothe:2010}.

The life-long interest of R.~Stora in the quantization of field theories based on the fundamental principle
of causality made him also contribute to general approaches as the one of Epstein and Glaser
(elaborating recently  on the problem of the extension  of distribution-valued field operators~\cite{Nikolov:2013nba}),
and work out what he considered to be the ``missing chapters'' of the subject~\cite{Stora:2009hc},
an endeavor that he could unfortunately not complete despite  relentless efforts and various successes.

While interested and knowledgeable in a vast spectrum of topics in physics, Raymond was always particularly
concerned with unveiling the underlying mathematical structures
or finding the appropriate mathematical framework for the formulation of theories or for the solution of problems,
e.g. anomalies, differential algebras, gauge fixing, cohomological field theories, \ldots~\cite{\kastlerstora}.
In all instances he was extremely attached to correct and precise statements
which generally contributed to clarify the issues, but also
retained him from publishing a certain number  of his results (leaving nicely hand-written
manuscripts communicated to friends and colleagues).
He shared his passion with various long term friends from the international ``Feldverein'' like
 C.~Itzykson,  B.~Zumino, D.~Kastler, A.S.~Wightman, J.~Wess,  H.-J.~Borchers, R.~Haag, \ldots, and it is very
sad to note that all of these masters left us fairly recently, their deep insights and precious advice being greatly missed.
As for Raymond, all of those who had the chance to meet him will always remember his culture, curiosity, enthusiasm,
his brilliant and penetrating insights,
as well as his great modesty, generosity and humanity.

\vskip 1.2truecm


\subsection*{\textbf{Acknowledgments}}



F.G. acknowledges stimulating discussions with F.~Delduc and H.~Samtleben
and wishes to thank M.~Dubois-Violette
and E.~Pilon for helpful comments.
We are grateful to J.-B.~Zuber for a careful reading of the text and for his pertinent remarks.
We also acknowledge the constructive feedback of S.~Deser and F.~Hehl concerning different points in the text.
Finally, we are indebted to the anonymous referee for his constructive and clarifying comments
which contributed to an improvement of the presentation.

\vskip 1.2truecm

\appendix

\section{Gauge field theories in a nutshell}
\label{sec:GFTnutshell}

To set the stage and fix the notation, we briefly recall the basics of non-Abelian gauge
theories\footnote{Non-Abelian gauge theories
have been discovered at about the same time and independently by   W.~Pauli, by R.~Shaw, by
C.~N.~Yang and R.~Mills and by R.~Utiyama who considered right away the case of general Lie groups.
However, the priority goes to Yang and Mills
who were the first to publish their results --- see reference~\cite{ORaifear} for the fascinating history of the subject.}.
 in this appendix
(e.g. see
references~\cite{Stora:1973dka,Stora:1976kd,Das:2008,Grensing:2013}).

\paragraph{General set-up:}
Let $G$ be a compact matrix Lie group (e.g. $G= SU({\cal N} )$) and $\lie$
its Lie algebra with basis elements
$\{ T^a \} _{a = 1,\dots,n_G}$ (with $n_G \equiv \txt{dim}\, G$) satisfying
\begin{equation}
[  T^a , T^b ]  =
\ri f^{abc}  T^c
\, .
\label{eq:LArelations}
\end{equation}
Here, the real structure constants $f^{abc}$ are totally antisymmetric in the indices\footnote{%
For semi-simple
Lie algebras like $SU({\cal N} )$,
the structure constants can be chosen to be totally antisymmetric.
Here, we recall~\cite{Frappat} that a real or complex Lie algebra is called \emph{semi-simple} if it does not contain any Abelian ideal
(i.e. invariant Lie subalgebra) except $\{ 0 \}$; the Lie algebra $\lie$ is called \emph{simple} if
it is not Abelian and does not contain any  ideals other than $\lie$ and $\{ 0 \}$.}.
Furthermore, we assume that the matrices $ T^a$ are
Hermitian, i.e. $(T^a) ^{\dagger} =  T^a$.
By virtue of relation~\eqref{eq:LArelations}, they are also traceless.
For instance, for $G=SU(2)$ we can choose
$T^a = \frac{1}{2} \, \sigma ^a$ where $\sigma ^1,\sigma ^2,\sigma ^3$ are the Pauli
matrices and the structure constants are then given by $\varepsilon^{abc}$.
 Since the indices $a,b,c$ are internal indices, they
can indifferently be written as upper or lower
indices, and Einstein's convention of summing over identical indices is also applied to them.

The element $\gro$ of the $n_G$-dimensional Lie group $G$ depends smoothly on $n_G$
real parameters $\omega^a$ and it can be written (at least in the vicinity of the identity of $G$)
as
\begin{align}
\label{eq:LG}
\gro = \re^{- \ri \mq {\omega}} \simeq \Id -  \ri \mq {\omega}
\, , \qquad \txt{with} \ \; {\omega} \equiv \omega^a  T_a
\, .
\end{align}
Here, the real constant $\mq$ represents the ``non-Abelian'' or ``YM charge''
(which could as well be absorbed in the parameters $\omega^a$ in this context)
and we have $\gro^\dagger = \gro^{-1}$.
The finite-dimensional \emph{structure group} $G$ gives rise to the infinite-dimensional
\emph{gauge group} ${\cal G} \equiv \{ \gro : \br^n \to G \}$ whose elements describe the gauge
transformations of matter and gauge fields which we will outline next.

\paragraph{Fields and transformation laws:}
We suppose that the \emph{matter} content is given by a multiplet $\vp \equiv [ \vp_1, \dots , \vp_N ]^t$,
i.e. a column vector with $N$ components $\vp_A$ each of which is a classical relativistic
field, e.g. we have a collection of complex scalar fields $\phi_A$ or of Dirac fields $\psi_A$.
For global gauge transformations, the multiplet $\vp$ is assumed to transform
with a $N$-dimensional unitary representation $\hg$ of the group $G$: this means that
the group element $\gro \in G$ acts on $\vp$ by a unitary $N\times N$ matrix $\hg$ satisfying
$\widehat{\gro_1 \gro_2} = \hg_1 \hg_2$.
This representation of $G$ is related to a $N$-dimensional representation $\hat T_a$ of the Lie algebra $\lie$
(see equation~\eqref{eq:LG}):
\[
\hg = \re^{- \ri \mq \hat{\omega}} \simeq \Id_N -  \ri \mq \hat{\omega}
\, , \qquad \txt{with} \quad \hat{\omega} \equiv \omega^a \hat T_a
\quad \txt{and} \quad \hat{T}_a^{\dagger} = \hat T _a
\, .
\]
For (local) gauge transformations, the symmetry parameters $\omega^a$
and the matrices $\hg$ depend smoothly on $x\in \br^n$.

The \emph{transformation law of the matter multiplet} $\vp$ and of its Hermitian conjugate
$\vp^\dagger$ read
\begin{align}
\label{eq:globgaugetr}
\preU\vp  = \hg \, \vp  \, ,
&
\qquad
\txt{i.e.} \quad \delta \vp = - \ri \mq \, \hat{\omega} \vp  \quad
\txt{or} \quad
 \delta \vp_A = - \ri \mq \, \hat{\omega}_{AB} \vp_B
 \,,
\nn\\
\preU\vp^\dagger  = \vp^\dagger  \hg^{-1} \, ,
&
\qquad
\txt{i.e.} \quad \delta \vp^\dagger =  \ri \mq \, \vp^\dagger \hat{\omega}  \quad
\txt{or} \quad
 \delta \vp^\dagger _A =  \ri \mq \, \vp^\dagger _B \hat{\omega}_{BA}
\, .
\end{align}
Here, we use the notation $\vp^\dagger _A \equiv (\vp^\dagger )_A $
and we note that the transformations only act on the
internal symmetry indices $A$ of the matter fields.

The \emph{covariant derivative of the matter multiplet} $\vp$ is defined by
\begin{align}
D_\m \vp &\equiv \pa_\m \vp + \ri \mq \hat A _\m \vp \, , &
D_\m \vp^\dagger &\equiv  (D_\m \vp)^\dagger = \pa_\m \vp^\dagger - \ri \mq \vp^\dagger \hat A _\m
\, .
\end{align}
Here, $\hat{A} _{\mu} (x) \equiv
A_{\mu} ^a (x) \hat T ^a $ is the $N$-dimensional representation
of the \emph{gauge potential}: the latter is a $\lie$-valued vector field ${A} _{\mu} (x) \equiv
A_{\mu} ^a (x) T^a $ where  $(A_{\mu} ^a)_{\m \in \{ 0, 1, \dots, n-1\}} $ is a real-valued
vector field  for each value of $a \in \{ 1, \dots , n_G \}$.
Under a finite gauge transformation $x\mapsto U(x)$, the gauge potential transforms
inhomogeneously with the \emph{adjoint representation} of the gauge group:
\begin{equation}
\label{eq:fgt}
\preU A_{\mu} = \gro A_{\mu} \gro^{-1}
- \frac{\ri}{\mq} \, \gro \pa_{\mu} \gro^{-1}
\, .
\end{equation}
For infinitesimal gauge variations $^{\gro \!\!} A_{\mu} \simeq A_\m + \delta A_\m$
it follows from~\eqref{eq:fgt} and~\eqref{eq:LG} that $A_\m$ transforms with the
\emph{covariant derivative of the $\lie$-valued functions} $x \mapsto \omega (x) \equiv \omega ^a(x) T_a$:
\begin{equation}
\label{eq:CDom}
\delta A_\m = D_\m \omega \equiv \pa_\m \omega  + \ri \mq \, [ A_\m , \omega ]
\, .
\end{equation}
This transformation law involves the non-Abelian charge $\mq$ (self-coupling constant)
which also appears in the matter transformation law~\eqref{eq:globgaugetr}.
For the Lie group generators $T_a$ appearing in $\omega \equiv \omega^a T_a$
and $A_\m \equiv A^a_\m T_a$, we can consider any representation $T_a \mapsto r(T_a)$
of the Lie algebra, e.g. $r(T_a) = \hat T _a$ as we did for the discussion of the matter multiplet $\vp$.

The commutator of two covariant derivatives determines the \emph{YM field strength tensor}
${F} _{\mu \nu} \equiv {F} _{\mu \nu}^a T_a$:
\begin{equation}
\label{eq:YMfs}
[ D_\m , D_\n ] \, \omega = \ri \mq \, [ F_{\m \n} , \omega ]
\, , \qquad \txt{with} \ \;
{F} _{\mu \nu} \equiv \pa_{\mu} {A} _{\nu} - \pa_{\nu}
{A} _{\mu} + \ri \mq \, [ {A} _{\mu}  ,  {A} _{\nu} ]
\, .
\end{equation}
Hence we have the gauge transformation law
\begin{equation}
\label{eq:YMftrans}
\preU{F} _{\mu \nu} = \gro {F} _{\mu \nu} \gro^{-1}
\qquad \txt{or} \qquad
 \delta {F} _{\mu \nu} =  \ri \mq \, [ {F}_{\mu \nu}, {\omega}  ]
\, ,
\end{equation}
i.e. $F_{\m \n}$ transforms with the adjoint representation.
The Jacobi identity for the covariant derivatives implies the \emph{Bianchi identity} for the
field strength:
$0=D_{\lambda} F_{\m \n} \, +$ cyclic permutations of the indices.
In contrast to the Abelian theory, the field strength presently involves terms which are non-linear
in the gauge field $A_\m$ and it is not invariant under gauge transformations.
Its components $F^a_{0i} \equiv E^a_{x^i}$, $F^a_{ij} = \varepsilon_{ijk} B^a_{x^k}$ (for $n=4$)
may be viewed as the non-Abelian generalization of the electric and magnetic fields of Maxwell's theory.

The theory can be generalized from Minkowski space
to an $n$-dimensional smooth manifold $M$ and a global (coordinate free) formulation can be given,
e.g. see reference~\cite{Bleecker:1981me}.
The latter relies on the introduction of a principal fiber bundle $P$ (with
compact Lie group $G$) over $M$,
i.e. a manifold $P$ which
has locally the structure $U \times G$ where $U$ is an open subset of $M$. The YM potential is then introduced
as a $\lie$-valued $1$-form on $P$ with appropriate transformation properties and it is referred to as a \emph{connection.}
The expressions given above for $\br^n$ then represent local expressions on $U\simeq \br^n $.

\paragraph{Dynamics:}
In the following, we generally drop the hat denoting the representation
(of the Lie algebra or Lie group) which is considered for the matter multiplets.
Concerning the normalization of the trace, we note that for
a simple Lie algebra one may choose a basis $(T_a)$ such that
$\Tr \, (r(T^a ) \, r(T^b) ) \, = \, c_{2} (r) \, \delta^{ab}$
for any representation $T_a \mapsto r(T_a)$,
where the constant $c_2(r)$ is known as the index of the representation $r$.
For instance, for the adjoint representation, $T_a$ is represented by
$(r(T_a))_{bc} = - \ri f_{abc}$ and $c_2 (r) =2$.

The dynamics of \emph{pure YM-theory} is described by the classical action
\begin{equation}
S_g [ A]  \equiv   -  \frac{1}{4 c_2} \, \int d^nx \ \Tr
\, ( {F} ^{\mu \nu} {F} _{\mu \nu} )
 = -
 \frac{1}{4}  \int d^nx
 \,{F} ^{a \mu \nu}
\, {F}^a _{\mu \nu}
 \, ,
 \label{eq:invact}
\end{equation}
which is invariant under gauge transformations.
Free Maxwell theory represents a particular case of pure YM theory
for which the gauge group $G= U(1)$ is Abelian: the internal index takes a single value
$a=1$ and the totally antisymmetric structure constants $f^{abc}$ in~\eqref{eq:LArelations} vanish,
as does the commutator term in the field strength~\eqref{eq:YMfs} and in the covariant derivative~\eqref{eq:CDom}.
Thus, in this particular case there is no self-interaction of gauge potentials in the action~\eqref{eq:invact}.

The dynamics of a matter multiplet  $\phi \equiv [ \phi_1, \dots , \phi_N ]^t$,
of $N$ scalar fields or of a multiplet $\psi \equiv [ \psi_1, \dots , \psi_N ]^t$
of $N$ Dirac fields of mass $m$ which are coupled to the gauge field $A_\m$ are respectively described by
the gauge invariant Lagrangian densities
\begin{align}
 {\cal L}_M (\phi , A ) \equiv (D^\mu \phi^\dagger) ( D_\mu \phi ) - m^2 \phi^\dagger \phi
 \, ,
 \label{eq:LagGM}
\end{align}
and
\begin{align}
{\cal L}_M (\psi, A) \equiv {\ri} \, \bar{\psi} \gamma^\m \! \stackrel{\leftrightarrow}{D}_{\mu} \!\! \psi - m \bar{\psi} {\psi}
\equiv
\frac{\ri}{2} \, \Big[ \bar{\psi} \gamma^\m {D}_{\mu} \psi  -  ({D}_{\mu} \bar{\psi} ) \gamma^\m \psi  \Big] - m \bar{\psi} {\psi}
 \, ,
 \label{eq:LMDiracYM}
\end{align}
where $\bar{\psi} \equiv [ \bar{\psi}_1, \dots , \bar{\psi}_N ]$
and $\bar{\psi}\gamma^\m \equiv [ \bar{\psi}_1 \gamma^\m , \dots , \bar{\psi}_N \gamma^\m ]$.
The equations of motion for interacting matter and gauge fields follow from the gauge invariant action functional
$S[\vp , A]  \equiv S_g [A] + S_M [\vp ,A]$: we have
\begin{align}
0 = \frac{\delta S}{ \delta \phi^\dagger} =  \frac{\delta S_M }{ \delta \phi^\dagger}
= - ( D^\mu D_\mu + m^2 ) \phi \, ,
\qquad \txt{or} \quad
0 = \frac{\delta S}{ \delta \bar{\psi}} =  \frac{\delta S_M}{\delta \bar{\psi}}
=  ( \ri \gamma^{\mu } D_\mu  - m) \psi
 \, ,
\label{eq:EOMmatterMul}
\end{align}
as well as the Hermitian conjugate expressions, and (for $a \in \{ 1, \dots , n_G \}$)
\begin{align}
0 = \frac{\delta S}{ \delta A^a _\mu} =  \frac{\delta S_g}{ \delta A^a_\mu} + \frac{\delta S_M}{ \delta A^a_\mu}
= D_\nu F_a^{\nu \mu} - j_a ^\mu
 \,, \qquad \txt{i.e.} \ \;
\Boxed{
D_\n F^{\n \m} = j^\m
}
\, ,
\label{eq:YMeq}
\end{align}
with
\begin{align}
\Boxed{
j_a ^\mu [\vp , A ] \equiv - \frac{\delta S_M [\vp, A]}{\delta A^a _\mu}
}
 \, .
 \label{eq:genNAmattercurrent}
\end{align}
From~\eqref{eq:genNAmattercurrent}, \eqref{eq:LagGM} and ~\eqref{eq:LMDiracYM},
we get the $\lie$-valued currents
\begin{align}
 j_a ^\mu [\phi , A ] &= \ri \mq \left[  \phi^\dagger \hat{T}_a D^\mu \phi
-  (D^{\mu } \phi^\dagger) \hat{T}_a  \phi \right]
\, , &
 j_a ^\mu [\psi , A ] &= \mq \, \bar{\psi}  \gamma^{\mu } \hat{T}_a \psi
 \, ,
\label{eq:matterNAcurrent}
\end{align}
which transform with the adjoint representation under gauge transformations:
\begin{align}
\label{eq:delJ}
\preU{j} ^{\mu} = \gro {j} ^{\mu} \gro^{-1}
\qquad \txt{or} \qquad
 \delta j^{\mu} = \ri \mq \, [ j^\m , {\omega}  ] \qquad \txt{with} \ \;
\omega \equiv \omega^a T_a \, .
\end{align}
The equation of motion~\eqref{eq:YMeq}
for the non-Abelian gauge field is referred to as the  \emph{Yang-Mills equation}.
In contrast to the Maxwell equation appearing in $U(1)$ gauge theory, the YM equation
involves the covariant derivative.
Hence the matter current $j^\mu [\vp , A ]$, which appears as a source in this equation, is only \emph{covariantly conserved}:
\begin{align}
D_\mu j^\mu = D_\mu D_\nu F^{\nu \mu} = \frac{1}{2} \, [ D_\mu , D_\nu ] F^{\nu \mu}
= -\frac{1}{2} \, \ri \mq \, [ F_{\mu \nu},  F^{\mu \nu} ] = 0\, ,
\qquad \txt{i.e.} \ \;
\Boxed{
D_\mu j^\mu = 0
}
\, .
\label{eq:CovConsLaw}
\end{align}



\providecommand{\href}[2]{#2}\begingroup\raggedright\endgroup


\begin{thebibliography}{100}
\small\itemsep=3pt
\tolerance 1414
\hbadness 1414
\emergencystretch 1.5em
\hfuzz 0.3pt
\widowpenalty=10000
\vfuzz \hfuzz
\raggedbottom

\bibitem{Weyl:1952}
H.~Weyl, \emph{Symmetry}, (Princeton Univ. Press, 1952).

\bibitem{Gieres:1997iw}
F.~Gieres, ``About symmetries in physics'', in \emph{Symmetries in Physics},
  F.~Gieres, M.~Kibler, C.~Lucchesi, and O.~Piguet, eds., (Editions
  Fronti{\`e}res, 1998).
\newblock
  \href{http://arxiv.org/abs/hep-th/9712154}{\texttt{arXiv:hep-th/9712154}}.
\newblock
Proceedings of the fifth S{\'e}minaire Rhodanien de Physique, Dolomieu 1997.
\newblock

\bibitem{Noether:1918zz}
E.~Noether, ``{Invariante Variationsprobleme}'',
\href{http://gdz.sub.uni-goettingen.de/dms/load/img/?PPN=GDZPPN00250510X}{\emph{Nachr. Ges. Wiss.
  G{\"o}tt.} \textbf{1918} (1918) 235--257},
  \href{http://arxiv.org/abs/physics/0503066}{\texttt{arXiv:physics/0503066
  [physics]}}.

\bibitem{Olver}
P.~J. Olver, \emph{{Applications of Lie Groups to Differential Equations}},
  second~ed., vol.~107 of \emph{Graduate Texts in Mathematics}, (Springer
  Verlag, 1993).

\bibitem{Kosmann}
Y.~Kosmann-Schwarzbach,
  \href{http://dx.doi.org/10.1007/978-0-387-87868-3}{\emph{{The Noether
  Theorems: Invariance and Conservation Laws in the Twentieth Century}}},
  Sources and Studies in the History of Mathematics and Physical Sciences,
  (Springer Verlag, 2011).

\bibitem{Sundermeyer:2014kha}
K.~Sundermeyer,
  \href{http://dx.doi.org/10.1007/978-94-007-7642-5}{\emph{{Symmetries in
  Fundamental Physics}}}, vol.~176 of \emph{Fundam. Theor. Phys.},
(Springer Verlag, 2014).
\newblock

\bibitem{Zuber:2013rha}
J.~B.~Zuber,
``{Invariances in physics and group theory}'',
Talk given at the Conference \emph{Lie and Klein: the Erlangen program and its
impact on mathematics and physics,  Strasbourg, Sept. 2012},
\href{https://arxiv.org/abs/1307.3970}{\texttt{arXiv:1307.3970 [hep-th]}}.

\bibitem{Banados:2016zim}
M.~Ba{\~{n}}ados and I.~A. Reyes, ``A short review on {Noether's} theorems,
  gauge symmetries and boundary terms, for students'',
\href{http://arxiv.org/abs/1601.03616}{\texttt{arXiv:1601.03616 [hep-th]}}.

\bibitem{Ogievetsky:1978js}
V.~Ogievetsky and E.~Sokatchev, ``Supercurrent'', \emph{Sov. J. Nucl. Phys.}
  \textbf{28} (1978) 423,
[Yad. Fiz. \textbf{28} (1978) 825].

\bibitem{Lang:1978ws}
W.~Lang, ``Currents in supersymmetric gauge theories'',
\href{http://dx.doi.org/10.1016/0550-3213(79)90300-6}{\emph{Nucl. Phys.}
  \textbf{B150} (1979) 201--220}.

\bibitem{Zahn:2003bt}
J.~Zahn, \href{http://dx.doi.org/10.3204/DESY-THESIS-2003-041}{``{Wirkungs- und
  Lokalit{\"a}tsprinzip f{\"u}r nichtkommutative skalare Feldtheorien}'',}
  Master's thesis, {Universit{\"a}t Hamburg}, 2003.

\bibitem{Ohanian:2013}
H.~C. Ohanian and R.~Ruffini,
  \href{http://dx.doi.org/10.1017/CBO9781139003391}{\emph{{Gravitation and
  Spacetime}}}, third~ed., (Cambridge Univ. Press, 2013).

\bibitem{Belinfante:1939}
F.~J. Belinfante, ``{On the spin angular momentum of mesons}'',
  \href{http://dx.doi.org/10.1016/S0031-8914(39)90090-X}{\emph{Physica}
  \textbf{6} (1939) 887--898}.

\bibitem{Belinfante:1940}
F.~J. Belinfante, ``{On the current and the density of the electric charge, the
  energy, the linear momentum and the angular momentum of arbitrary fields}'',
  \href{http://dx.doi.org/10.1016/S0031-8914(40)90091-X}{\emph{Physica}
  \textbf{7} (1940) 449}.


\bibitem{Grensing:2013}
G.~Grensing, \href{http://dx.doi.org/10.1142/8771}{\emph{Structural Aspects of
  Quantum Field Theory and Noncommutative Geometry}}, (World Scientific Publ.,
  2013).

\bibitem{Ryder:1996}
L.~H. Ryder, \emph{Quantum Field Theory}, second~ed., (Cambridge Univ. Press,
  1996).

\bibitem{Forger:2003ut}
M.~Forger and H.~R{\"o}mer, ``{Currents and the energy momentum tensor in
  classical field theory: A fresh look at an old problem}'',
  \href{http://dx.doi.org/10.1016/j.aop.2003.08.011}{\emph{Annals Phys.}
  \textbf{309} (2004) 306--389},
\href{http://arxiv.org/abs/hep-th/0307199}{\texttt{arXiv:hep-th/0307199}}.

\bibitem{GotayMarsden}
M.~J. Gotay and J.~E. Marsden,
``{Stress-energy-momentum tensors and the Belinfante-Rosenfeld formula}'',
\href{http://www.cds.caltech.edu/~marsden/bib/1992/05-GoMa1992/}{\emph{Contemp. Math.} \textbf{132} (1992) 367-392},

\bibitem{LandauLifschitz2}
L.~Landau and E.~Lifshitz,
  \href{http://dx.doi.org/10.1016/B978-0-08-025072-4.50001-0}{\emph{The
  Classical Theory of Fields, Course of Theoretical Physics, Vol. 2}},
  fourth~ed., (Butterworth-Heinemann, 1980).

\bibitem{Hanson:1976cn}
A.~J. Hanson, T.~Regge, and C.~Teitelboim,
\href{http://hdl.handle.net/2022/3108}{\emph{Constrained Hamiltonian
  Systems}}, (Roma: Accademia Nazionale dei Lincei, 1976).
\newblock

\bibitem{IP}
D.~N. Blaschke, F.~Gieres, M.~Reboud and M.~Schweda,
``{Poincar\'e transformations in the Hamiltonian formulation of gauge field theories}'',
talk presented by F.~Gieres at the XXIVth International Conference on \emph{Integrable Systems and Quantum symmetries} (ISQS-24),
Prague, June 2016, manuscript in preparation.



\bibitem{Duncan:2012}
A.~Duncan, \emph{The Conceptual Framework of Quantum Field Theory}, (Oxford
  Univ. Press, 2012).

\bibitem{Griffiths:2012}
D.~J. Griffiths, ``{Resource letter EM-1: Electromagnetic momentum}'',
  \href{http://dx.doi.org/10.1119/1.3641979}{\emph{Am. J. Phys.} \textbf{80}
  (2012) 7--18}.

\bibitem{Schroder:1990ri}
U.~E. Schr{\"o}der, \href{http://dx.doi.org/10.1142/7689}{\emph{{Special
  Relativity}}}, vol.~33 of \emph{World Sci. Lect. Notes Phys.},
(Singapore: World Scientific, 1990).
\newblock

\bibitem{Itzykson:2005}
C.~Itzykson and J.-B. Zuber, \emph{Quantum Field Theory}, {Dover}~ed., (New
  York: Dover Publ. Inc., 2005).

\bibitem{Kosyakov:2007qc}
B.~P. Kosyakov, \emph{{Introduction to the Classical Theory of Particles and
  Fields}},
(Springer Verlag, 2007).
\newblock

\bibitem{GellMann:1960np}
M.~Gell-Mann and M.~L{\'e}vy, ``{The axial vector current in beta decay}'',
\href{http://dx.doi.org/10.1007/BF02859738}{\emph{Nuovo Cim.} \textbf{16}
  (1960) 705}.

\bibitem{Callan:1970ze}
C.~G. Callan~Jr., S.~R. Coleman, and R.~Jackiw, ``{A new improved
  energy-momentum tensor}'',
\href{http://dx.doi.org/10.1016/0003-4916(70)90394-5}{\emph{Annals Phys.}
  \textbf{59} (1970) 42--73}.

\bibitem{Jackiw:2011vz}
R.~Jackiw and S.-Y. Pi, ``Tutorial on scale and conformal symmetries in diverse
  dimensions'',
  \href{http://dx.doi.org/10.1088/1751-8113/44/22/223001}{\emph{J. Phys. A:
  Math. Theor.} \textbf{44} (2011) 223001},
\href{http://arxiv.org/abs/1101.4886}{\texttt{arXiv:1101.4886 [math-ph]}}.

\bibitem{DiFrancesco:1997nk}
P.~Di~Francesco, P.~Mathieu, and D.~S{\'{e}}n{\'{e}}chal,
  \href{http://dx.doi.org/10.1007/978-1-4612-2256-9}{\emph{{Conformal Field
  Theory}}}, {Graduate Texts in Contemporary Physics},
(New York: Springer, 1997).
\newblock

\bibitem{Thirring:1997}
W.~Thirring, \href{http://dx.doi.org/10.1007/978-1-4612-0681-1}{\emph{Classical
  Mathematical Physics: Dynamical Systems and Field Theories}}, third~ed.,
  (Springer Verlag, 1997).

\bibitem{Sexl:2001}
U.~Sexl and H.~K. Urbantke,
  \href{http://dx.doi.org/10.1007/978-3-7091-6234-7}{\emph{Relativity, Groups,
  Particles: Special Relativity and Relativistic Symmetry in Field and Particle
  Physics}}, (Springer Verlag, 2001).

\bibitem{Hawking:1973uf}
S.~W. Hawking and G.~F.~R. Ellis,
  \href{http://dx.doi.org/10.1017/CBO9780511524646}{\emph{The Large Scale
  Structure of Space-time}}, Cambridge Monographs on Mathematical Physics,
  (Cambridge Univ. Press, 1973).

\bibitem{Senovilla:2014gza}
J.~M.~M. Senovilla and D.~Garfinkle, ``{The 1965 Penrose singularity
  theorem}'',
  \href{http://dx.doi.org/10.1088/0264-9381/32/12/124008}{\emph{Class. Quant.
  Grav.} \textbf{32} (2015) 124008},
\href{http://arxiv.org/abs/1410.5226}{\texttt{arXiv:1410.5226 [gr-qc]}}.

\bibitem{Wald:1984}
R.~M. Wald, \emph{{General Relativity}}, (Chicago: Univ. Press, 1984).

\bibitem{Poisson-book}
E.~Poisson, \emph{A Relativist's Toolkit: The Mathematics of Black-Hole
  Mechanics}, (Cambridge Univ. Press, 2008).

\bibitem{Eriksen:1979vq}
E.~Eriksen and J.~M. Leinaas, ``Gauge invariance and the transformation
  properties of the electromagnetic four-potential'',
\href{http://dx.doi.org/10.1088/0031-8949/22/3/003}{\emph{Phys. Scripta}
  \textbf{22} (1980) 199}.

\bibitem{Takahashi:1985dt}
Y.~Takahashi, ``Energy-momentum tensors in relativistic and non-relativistic
  classical field theory'',
  \href{http://dx.doi.org/10.1002/prop.19860340503}{\emph{Fortschr. Phys.}
  \textbf{34} (1986) 323--344}.

\bibitem{Munoz:1996wp}
G.~Mu{\~{n}}oz, ``{Lagrangian field theories and energy-momentum tensors}'',
\href{http://dx.doi.org/10.1119/1.18336}{\emph{Am. J. Phys.} \textbf{64} (1996)
  1153--1157}.

\bibitem{montesinos}
M.~Montesinos and E.~Flores, ``{Symmetric energy-momentum tensor in Maxwell,
  Yang-Mills, and Proca theories obtained using only Noether's theorem}'',
\href{http://rmf.smf.mx/pdf/rmf/52/1/52_1_29.pdf}{\emph{Rev. Mex. Fis.} \textbf{52} (2006) 29--36},
  \href{http://arxiv.org/abs/hep-th/0602190}{\texttt{arXiv:hep-th/0602190}}.

R.~E. Gamboa~Saravi, ``{On the energy momentum tensor}'',
  \href{http://dx.doi.org/10.1088/0305-4470/37/40/017}{\emph{J. Phys. A: Math.
  Gen.} \textbf{37} (2004) 9573--9586},
\href{http://arxiv.org/abs/math-ph/0306020}{\texttt{arXiv:math-ph/0306020}}.

\bibitem{Barut:1981}
A.~O. Barut, \emph{{Electrodynamics and Classical Theory of Fields and
  Particles}}, {Dover Books on Physics}, (Dover Publ. Inc., 1981).

\bibitem{Rosenfeld:1940}
L.~Rosenfeld, ``{Sur le tenseur d'impulsion-\'energie}'', \href{http://www2.academieroyale.be/academie/documents/XVIII_6_RosenfeldL._Su%
rletenseurdimpulsionenergie_194019293.pdf}{\emph{M{\'e}moires
  Acad. Roy. de Belgique} \textbf{18} (1940) 1--30}.

\bibitem{Blagojevic:2013xpa}
M.~Blagojevi{\'c} and F.~W. Hehl, eds., \href{http://www.worldscientific.com/worldscibooks/10.1142/p781}{\emph{Gauge Theories of Gravitation: A
  Reader with Commentaries}}, (World Scientific, 2013).


\bibitem{hehlhammond}
F.~W. Hehl, P.~von~der Heyde, G.~D. Kerlick, and J.~M. Nester, ``General
  relativity with spin and torsion: {Foundations} and prospects'',
\href{http://dx.doi.org/10.1103/RevModPhys.48.393}{\emph{Rev. Mod. Phys.}
  \textbf{48} (1976) 393--416}.

R.~T. Hammond, ``{Torsion gravity}'',
\href{http://dx.doi.org/10.1088/0034-4885/65/5/201}{\emph{Rept. Prog. Phys.}
  \textbf{65} (2002) 599--649}.

\bibitem{TrautmanEMP}
A.~Trautman, ``{Einstein-Cartan theory}'', in \emph{Encyclopedia of
  Mathematical Physics}, J.-P. Fran{\c c}oise, G.~L. Naber, and S.~T. Tsou,
  eds., vol.~2, pp.~189--195, (Oxford: Elsevier, 2006).
\newblock
\href{http://arxiv.org/abs/gr-qc/0606062}{\texttt{arXiv:gr-qc/0606062}}.
\newblock

\bibitem{Hehl:2014eja}
F.~W. Hehl, ``On energy-momentum and spin/helicity of quark and gluon fields'',
  in \emph{{Proceedings, 15th Workshop on High Energy Spin Physics (DSPIN-13):
  Dubna, Russia, Oct 8-12, 2013}}, (Dubna: JINR, 2014).
\newblock
\href{http://arxiv.org/abs/1402.0261}{\texttt{arXiv:1402.0261 [gr-qc]}}.
\newblock

\bibitem{Streater:1989vi}
R.~F. Streater and A.~S. Wightman, \emph{{PCT, spin and statistics, and all
  that}}, {Princeton Landmarks in Mathematics and Physics},
(Princeton Univ. Press, 2000).
\newblock

\bibitem{Haag:1992hx}
R.~Haag, \href{http://dx.doi.org/10.1007/978-3-642-61458-3}{\emph{Local Quantum
  Physics: Fields, Particles, Algebras}}, second~ed., {Texts and Monographs in
  Physics},
(Berlin: Springer Verlag, 1996).
\newblock

\bibitem{Jackson:1998}
J.~D. Jackson, \emph{Classical Electrodynamics}, third~ed., (Wiley, 1998).

\bibitem{Felsager:1981iy}
B.~Felsager,
  \href{http://dx.doi.org/10.1007/978-1-4612-0631-6}{\emph{{Geometry, Particles
  and Fields}}}, {Graduate Texts in Contemporary Physics},
(Springer Verlag, 1998).
\newblock

\bibitem{Utiyama:1959}
R.~Utiyama, ``Theory of invariant variation and the generalized canonical
  dynamics'', \href{http://dx.doi.org/10.1143/PTPS.9.19}{\emph{Prog. Theor.
  Phys. Suppl.} \textbf{9} (1959) 19--44}.

\bibitem{Balasin:2014dma}
H.~Balasin, D.~N. Blaschke, F.~Gieres, and M.~Schweda, ``{Wong's equations and
  charged relativistic particles in non-commutative space}'',
  \href{http://dx.doi.org/10.3842/SIGMA.2014.099}{\emph{SIGMA} \textbf{10}
  (2014) 099},
\href{http://arxiv.org/abs/1403.0255}{\texttt{arXiv:1403.0255 [hep-th]}}.


\bibitem{Deser:1970hs}
S.~Deser, ``Scale invariance and gravitational coupling'',
\href{http://dx.doi.org/10.1016/0003-4916(70)90402-1}{\emph{Annals Phys.}
  \textbf{59} (1970) 248--253}.

\bibitem{deser}
S.~Deser, ``{Self-interaction and gauge invariance}'',
  \href{http://dx.doi.org/10.1007/BF00759198}{\emph{Gen. Rel. Grav.} \textbf{1}
  (1970) 9--18},
\href{http://arxiv.org/abs/gr-qc/0411023}{\texttt{arXiv:gr-qc/0411023}}.

S.~Deser, ``{Gravity from self-interaction redux}'',
  \href{http://dx.doi.org/10.1007/s10714-009-0912-9}{\emph{Gen. Rel. Grav.}
  \textbf{42} (2010) 641--646},
  \href{http://arxiv.org/abs/0910.2975}{\texttt{arXiv:0910.2975 [gr-qc]}}.

\bibitem{Ortin}
T.~Ort{\'{\i}}n, \emph{{Gravity and Strings}}, second~ed., Cambridge Monographs
  on Mathematical Physics, (Cambridge Univ. Press, 2015).

\bibitem{Feynman-book}
R.~P. Feynman, F.~B. Morinigo, and W.~G. Wagner, \emph{{Feynman Lectures On
  Gravitation}}, Frontiers in Physics,
B.~Hatfield ed., (Westview Press, 2002).

\bibitem{straumann}
N.~Straumann, \emph{{General Relativity}}, second~ed., Graduate Texts in
  Physics, (Springer Verlag, 2013).

N.~Straumann, ``{Reflections on gravity}'',
  Concluding talk at the ESA-CERN Workshop, CERN, 5-7 April 2000,
\href{http://arxiv.org/abs/astro-ph/0006423}{\texttt{arXiv:astro-ph/0006423}}.

\bibitem{YM50}
 G.~'t~Hooft (ed.),
  \emph{{50 years of Yang-Mills theory}},
 (World Scientific Publ., 2005).

\bibitem{Birrell:1982ix}
N.~D. Birrell and P.~C.~W. Davies,
  \href{http://dx.doi.org/10.1017/CBO9780511622632}{\emph{{Quantum Fields in
  Curved Space}}}, Cambridge Monogr. Math. Phys.,
(Cambridge Univ. Press, 1982).
\newblock

\bibitem{Misner:1973}
C.~W. Misner, K.~S. Thorne, and J.~A. Wheeler, \emph{Gravitation}, (New York:
  W. H. Freeman and Company, 1973).

\bibitem{Hilbert:1915tx}
D.~Hilbert, ``{Die Grundlagen der Physik. (Erste Mitteilung.)}'', \href{http://gdz.sub.uni-goettingen.de/dms/load/img/?PPN=GDZPPN002504286}{\emph{Nachr.
  d. K. Gesellsch. d. Wiss. zu G{\"o}ttingen, Math.-Phys. Klasse} \textbf{1915,
  Heft 3} (1915) 395--407}.

\bibitem{einstein}
A.~Einstein, ``{Feldgleichungen der Gravitation}'',
\href{http://adsabs.harvard.edu/abs/1915SPAW.......844E}{\emph{Preuss. Akad. Wiss.,
  Sitzungsberichte} \textbf{1915 (Teil 2)} (1915) 844--847}.

A.~Einstein, ``{Die Grundlage der allgemeinen Relativit{\"a}tstheorie}'',
\href{http://dx.doi.org/10.1002/andp.200590044}{\emph{Annalen Phys.}
  \textbf{49} (1916) 769--822}.

\bibitem{Abbott:1981ff}
L.~F. Abbott and S.~Deser, ``{Stability of Gravity with a Cosmological
  Constant}'',
\href{http://dx.doi.org/10.1016/0550-3213(82)90049-9}{\emph{Nucl. Phys.}
  \textbf{B195} (1982) 76}.

\bibitem{DuboisViolette:1986ws}
M.~Dubois-Violette and J.~Madore, ``Conservation laws and integrability
  conditions for gravitational and {Yang-Mills} field equations'',
\href{http://dx.doi.org/10.1007/BF01210612}{\emph{Commun. Math. Phys.}
  \textbf{108} (1987) 213}.

\bibitem{Barnich:2001jy}
G.~Barnich and F.~Brandt, ``Covariant theory of asymptotic symmetries,
  conservation laws and central charges'',
  \href{http://dx.doi.org/10.1016/S0550-3213(02)00251-1}{\emph{Nucl. Phys.}
  \textbf{B633} (2002) 3--82},
\href{http://arxiv.org/abs/hep-th/0111246}{\texttt{arXiv:hep-th/0111246}}.

\bibitem{StefaniGR}
H.~Stephani, \emph{{Relativity --- An Introduction to Special and General
  Relativity}}, third~ed., (Cambridge Univ. Press, 2006).

\bibitem{ORaifear}
L.~O'Raifeartaigh, \emph{{The Dawning of Gauge Theory}}, Princeton Series in
  Physics, (Princeton Univ. Press, 1997).

\bibitem{Bertlmann-book}
R.~A. Bertlmann,
  \href{http://dx.doi.org/10.1093/acprof:oso/9780198507628.001.0001}{\emph{{Anomalies
  in Quantum Field Theory}}}, vol.~91 of \emph{{International Series of
  Monographs on Physics}},
(Oxford, UK: Clarendon, 1996).
\newblock

\bibitem{Rouet:1972ut}
A.~Rouet and R.~Stora, ``{Ward identities and trace identities for the
  stress-energy tensor}'',
\href{http://dx.doi.org/10.1007/BF02907133}{\emph{Lett. Nuovo Cim.} \textbf{4}
  (1972) 136--138}.

\bibitem{Rouet:1972bis}
A.~Rouet and R.~Stora, ``{Noether's theorem and Ward identities in
  renormalizable theories}'',
  \href{http://dx.doi.org/10.1007/BF02907134}{\emph{Lett. Nuovo Cim.}
  \textbf{4} (1972) 139}.

\bibitem{Stora:1973dka}
R.~Stora, ``Lagrangian field theory'', in \emph{Particle Physics}, Proceedings of the Les Houches Summer School on Theoretical Physics 1971,
C.~DeWitt
  and C.~Itzykson, eds., pp.~1--80, (Gordon and Breach, 1973).

\bibitem{Srednicki:2007qs}
M.~Srednicki, \emph{Quantum Field Theory},
(Cambridge Univ. Press, 2007).
\newblock

\bibitem{Duetsch:2001sw}
M.~D{\"u}tsch and F.~M. Boas, ``{The Master Ward identity}'',
  \href{http://dx.doi.org/10.1142/S0129055X02001454}{\emph{Rev. Math. Phys.}
  \textbf{14} (2002) 977},
\href{http://arxiv.org/abs/hep-th/0111101}{\texttt{arXiv:hep-th/0111101}}.

\bibitem{Avery:2015rga}
S.~G. Avery and B.~U.~W. Schwab, ``{Noether's second theorem and Ward
  identities for gauge symmetries}'',
  \href{http://dx.doi.org/10.1007/JHEP02(2016)031}{\emph{JHEP} \textbf{02}
  (2016) 031},
\href{http://arxiv.org/abs/1510.07038}{\texttt{arXiv:1510.07038 [hep-th]}}.

\bibitem{Coleman-Lectures}
S.~Coleman, \href{http://dx.doi.org/10.1017/CBO9780511565045}{\emph{Aspects of
  Symmetry}}, (Cambridge Univ. Press, 1985).

\bibitem{Fulling:1989nb}
S.~A. Fulling, \emph{Aspects of Quantum Field Theory in Curved Space-time},
  vol.~17 of \emph{London Math. Soc. Student Texts},
(Cambridge Univ. Press, 1989).
\newblock

\bibitem{Duff:1993wm}
M.~J. Duff, ``{Twenty years of the Weyl anomaly}'',
  \href{http://dx.doi.org/10.1088/0264-9381/11/6/004}{\emph{Class. Quant.
  Grav.} \textbf{11} (1994) 1387--1404},
\href{http://arxiv.org/abs/hep-th/9308075}{\texttt{arXiv:hep-th/9308075
  [hep-th]}}.

\bibitem{Blumenhagen:2013fgp}
R.~Blumenhagen, D.~L{\"u}st, and S.~Theisen,
  \href{http://dx.doi.org/10.1007/978-3-642-29497-6}{\emph{Basic Concepts of
  String Theory}}, Theoretical and Mathematical Physics,
(Springer Verlag, 2013).
\newblock


\bibitem{dymarskynakayama}
Y.~Nakayama, ``{A lecture note on scale invariance vs conformal invariance}'',
  \href{http://dx.doi.org/doi:10.1016/j.physrep.2014.12.003}{\emph{Phys. Rept.}
  \textbf{569} (2014) 1--93},
  \href{http://arxiv.org/abs/1302.0884}{\texttt{arXiv:1302.0884 [hep-th]}},
Lectures given at Taiwan Central University for the 5th Taiwan School on
  Strings and Fields.

A.~Dymarsky, Z.~Komargodski, A.~Schwimmer, and S.~Theisen, ``On scale and
  conformal invariance in four dimensions'',
  \href{http://dx.doi.org/10.1007/JHEP10(2015)171}{\emph{JHEP} \textbf{10}
  (2015) 171},
\href{http://arxiv.org/abs/1309.2921}{\texttt{arXiv:1309.2921 [hep-th]}}.

\bibitem{Blaschke:2014ioa}
D.~N. Blaschke, R.~Carballo-Rubio, and E.~Mottola, ``Fermion pairing and the
  scalar boson of the {2D} conformal anomaly'',
  \href{http://dx.doi.org/10.1007/JHEP12(2014)153}{\emph{JHEP} \textbf{12}
  (2014) 153},
\href{http://arxiv.org/abs/1407.8523}{\texttt{arXiv:1407.8523 [hep-th]}}.




\bibitem{becchi}
C.~Becchi, A.~Rouet, and R.~Stora, ``{The Abelian Higgs-Kibble model: Unitarity
  of the S operator}'',
\href{http://dx.doi.org/10.1016/0370-2693(74)90058-6}{\emph{Phys. Lett.}
  \textbf{B52} (1974) 344}.

C.~Becchi, A.~Rouet, and R.~Stora, ``Renormalization of the {Abelian
  Higgs-Kibble} model'',
\href{http://dx.doi.org/10.1007/BF01614158}{\emph{Commun. Math. Phys.}
  \textbf{42} (1975) 127--162}.

\bibitem{Becchi:1975nq}
C.~Becchi, A.~Rouet, and R.~Stora, ``Renormalization of gauge theories'',
\href{http://dx.doi.org/10.1016/0003-4916(76)90156-1}{\emph{Annals Phys.}
  \textbf{98} (1976) 287--321}.

\bibitem{Tyutin:1975qk}
I.~V. Tyutin, ``Gauge invariance in field theory and statistical physics in
  operator formalism'',
  \href{http://arxiv.org/abs/0812.0580}{\texttt{arXiv:0812.0580 [hep-th]}},
(Lebedev Institute preprint in Russian 1975: LEBEDEV-75-39).

\bibitem{Becchi:1975dw}
C.~Becchi, A.~Rouet, and R.~Stora,
  \href{http://dx.doi.org/10.1007/978-94-010-1490-8_9}{``Gauge field models'',}
  in \emph{{Renormalization Theory}},
  {Proceedings of the NATO Advanced Study Institute, Erice, August 1975},
  G.~Velo and A.~S. Wightman, eds., vol.~23
  of \emph{{NATO Advanced Study Institutes Series, Series C}}, pp.~269--297,
  (Dortrecht: Springer, 1976).
\newblock

\bibitem{Stora:1976kd}
R.~Stora, \href{http://dx.doi.org/10.1007/978-1-4615-8918-1_8}{``Continuum
  gauge theories'',} in \emph{{New Developments in Quantum Field Theory and
  Statistical Mechanics Carg{\`e}se 1976}},
  {XVI Latin American School of Physics Caracas, Venezuela 1976},
  M.~L{\'e}vy and P.~Mitter, eds.,
  vol.~26 of \emph{{NATO Advanced Study Institutes Series}}, pp.~201--224,
  (Springer US, 1977).
\newblock

\bibitem{Becchi:1981jx}
C.~Becchi, A.~Rouet, and R.~Stora,
  \href{http://dx.doi.org/10.1007/978-94-009-8368-7}{``Renormalizable theories
  with symmetry breaking'',} in \emph{{Field Theory, Quantization and
  Statistical Physics}}, E.~Tirapegui, ed., vol.~6 of \emph{Mathematical
  Physics and Applied Mathematics},
(Springer Verlag, 1981).
\newblock

\bibitem{Piguet:1980nr}
O.~Piguet and A.~Rouet, ``Symmetries in perturbative quantum field theory'',
\href{http://dx.doi.org/10.1016/0370-1573(81)90066-1}{\emph{Phys. Rept.}
  \textbf{76} (1981) 1}.

\bibitem{Becchi:1985bd}
C.~Becchi, ``Lectures on the renormalization of gauge theories'', in
  \emph{{Relativity, Groups and Topology}},
  {Les Houches Summer School on Theoretical Physics, August 1983},
  B.~DeWitt and R.~Stora, eds.,
  (Amsterdam: North-Holland, 1984).
\newblock

\bibitem{Piguet:1995}
O.~Piguet and S.~P. Sorella,
  \href{http://dx.doi.org/10.1007/978-3-540-49192-7}{\emph{Algebraic
  Renormalization: Perturbative Renormalization, Symmetries and Anomalies}},
  vol.~28 of \emph{Lect. Notes Phys.: Monographs},
(Springer Verlag, 1995).
\newblock

\bibitem{stora-1980ies}
R.~Stora, \href{http://dx.doi.org/10.1007/BFb0033731}{``Remarks on Slavnov
  symmetries'',} in \emph{Renormalization of Quantum Field Theories with
  Non-linear Field Transformations},
  Proceedings of a workshop at Ringberg Castle Tegernsee, 1987,
  P.~Breitenlohner, D.~Maison, and
  K.~Sibold, eds., vol.~303 of \emph{Lecture Notes in Physics}, pp.~213--219,
  (Springer Verlag, 1987).
\newblock

R.~Stora, ``{The Slavnov symmetry, cousins and descendants}'', in \emph{{BRS
  symmetry}},
  Proceedings of the International Symposium on the Occasion of its 20th
  Anniversary, Kyoto, Sept. 1995,
  M.~Abe, N.~Nakanishi, and I.~Ojima, eds., pp.~1--16, (Tokyo,
  Japan: Univ. Acad. Press, 1996).
\newblock

\bibitem{FrancoPiguet}
D.~H.~T. Franco and O.~Piguet, ``{A}lgebraic renormalization'',
  \href{http://dx.doi.org/10.4249/scholarpedia.8336}{\emph{Scholarpedia}
  \textbf{8} no.~11, (2013) 8336}.
  \url{http://www.scholarpedia.org/article/Algebraic_renormalization}.

\bibitem{Schweda-book:1998}
A.~Boresch, S.~Emery, O.~Moritsch, M.~Schweda, T.~Sommer, and H.~Zerrouki,
  \href{http://dx.doi.org/10.1142/3778}{\emph{Applications of Noncovariant
  Gauges in the Algebraic Renormalization Procedure}}, (World
  Scientific, 1998).

\bibitem{Wess:1971yu}
J.~Wess and B.~Zumino, ``{Consequences of anomalous Ward identities}'',
\href{http://dx.doi.org/10.1016/0370-2693(71)90582-X}{\emph{Phys. Lett.}
  \textbf{B37} (1971) 95--97}.

\bibitem{Zumino:1983ew}
B.~Zumino, ``Chiral anomalies and differential geometry'', in \emph{Relativity, Groups and Topology},
  {Les Houches Summer School on Theoretical Physics, August 1983},
  B.~DeWitt and R.~Stora, eds., (Amsterdam:
  North-Holland, 1984).
\newblock
\url{http://www.osti.gov/energycitations/product.biblio.jsp?osti_id=5461614}.

\bibitem{Stora:1983ct}
R.~Stora, \href{http://dx.doi.org/10.1007/978-1-4757-0280-4_19}{``Algebraic
  structure and topological origin of anomalies'',} in \emph{{Progress in Gauge
  Field Theory}}, {Carg{\`e}se Summer Institute, Sept. 1983},
  G.~'t~Hooft, A.~Jaffe, H.~Lehmann, P.~K. Mitter, I.~M.
  Singer, and R.~Stora, eds., vol.~115 of \emph{{NATO Advanced Study Institutes
  Series, Series B}}, pp.~543--562, (Plenum Press, 1984).
\newblock

\bibitem{Manes:1985df}
J.~Ma\~{n}es, R.~Stora, and B.~Zumino, ``Algebraic study of chiral anomalies'',
\href{http://dx.doi.org/10.1007/BF01208825}{\emph{Commun. Math. Phys.}
  \textbf{102} (1985) 157}.

\bibitem{Langouche:1984gn}
F.~Langouche, T.~Sch{\"u}cker, and R.~Stora, ``Gravitational anomalies of the
  adler-bardeen type'',
\href{http://dx.doi.org/10.1016/0370-2693(84)90057-1}{\emph{Phys. Lett.}
  \textbf{B145} (1984) 342--346}.

\bibitem{GirardiFerrara}
G.~Girardi, R.~Grimm, and R.~Stora, ``Chiral anomalies in $n=1$ supersymmetric
  {Yang-Mills} theories'',
\href{http://dx.doi.org/10.1016/0370-2693(85)91510-2}{\emph{Phys. Lett.}
  \textbf{B156} (1985) 203--208}.

S.~Ferrara, A.~Masiero, M.~Porrati, and R.~Stora, ``{Bardeen anomaly and
  Wess-Zumino term in the supersymmetric standard model}'',
  \href{http://dx.doi.org/10.1016/0550-3213(94)90545-2}{\emph{Nucl. Phys.}
  \textbf{B417} (1994) 238--256},
\href{http://arxiv.org/abs/hep-th/9311038}{\texttt{arXiv:hep-th/9311038}}.

\bibitem{Baulieu:1986hw}
L.~Baulieu, C.~Becchi, and R.~Stora, ``On the covariant quantization of the
  free bosonic string'',
\href{http://dx.doi.org/10.1016/0370-2693(86)90134-6}{\emph{Phys. Lett.}
  \textbf{B180} (1986) 55--60}.

\bibitem{lazzarini}
S.~Lazzarini and R.~Stora,
  \href{http://dx.doi.org/10.1007/978-94-011-7976-8_8}{``Ward identities for
  conformal models'',} in \emph{Stochastics, Algebra and Analysis in Classical
  and Quantum Dynamics}, S.~Albeverio, {\relax Ph}.~Blanchard, and D.~Testard,
  eds.,
  {Proceedings of the IVth French-German Encounter on Mathematics and Physics,
  CIRM, Marseille, France, February/March 1988},
  vol.~59 of \emph{Mathematics and Its Applications}, pp.~169--172,
  (Kluwer Academic Publishers, 1990).

S.~Lazzarini and R.~Stora,
  ``\href{http://www.iaea.org/inis/collection/NCLCollectionStore/_Public/22/013/22013087.pdf}{Ward identities for Lagrangian conformal
  models}'', in \emph{Knots, topology and quantum field theories},
  {Proceedings of the Johns Hopkins Workshop on Current Problems in Particle
  Theory 13, Florence, 1989},
  L.~Lusanna, ed., (World Scientific Publ., 1989).
\newblock

\bibitem{Ouvry:1988mm}
S.~Ouvry, R.~Stora, and P.~van Baal, ``On the algebraic characterization of
  {Witten's} topological {Yang-Mills} theory'',
\href{http://dx.doi.org/10.1016/0370-2693(89)90029-4}{\emph{Phys. Lett.}
  \textbf{B220} (1989) 159}.

\bibitem{Henneaux:1992}
M.~Henneaux and C.~Teitelboim, \emph{{Quantization of Gauge Systems}},
  (Princeton Univ. Press, 1992).


\bibitem{Gitman:1990}
D.~M. Gitman and I.~V. Tyutin,
  \href{http://dx.doi.org/10.1007/978-3-642-83938-2}{\emph{{Quantization of
  Fields with Constraints}}}, Springer Series in Nuclear and Particle Physics,
  (Springer Verlag, 1990).

\bibitem{Rothe:2010}
H.~J. Rothe and K.~D. Rothe,
  \href{http://dx.doi.org/10.1142/7689}{\emph{{Classical and Quantum Dynamics
  of Constrained Hamiltonian Systems}}}, vol.~81 of \emph{World Sci. Lect.
  Notes Phys.}, (World Scientific Publ., 2010).

\bibitem{Nikolov:2013nba}
N.~M. Nikolov, R.~Stora, and I.~Todorov, ``Renormalization of massless
  {Feynman} amplitudes in configuration space'',
  \href{http://dx.doi.org/10.1142/S0129055X14300027}{\emph{Rev. Math. Phys.}
  \textbf{26} (2014) 1430002},
\href{http://arxiv.org/abs/1307.6854}{\texttt{arXiv:1307.6854 [hep-th]}}.

\bibitem{Stora:2009hc}
R.~Stora, ``Renormalized perturbation theory: A missing chapter'',
  \href{http://dx.doi.org/10.1142/S0219887808003363}{\emph{Int. J. Geom. Meth.
  Mod. Phys.} \textbf{5} (2008) 1345--1360},
\href{http://arxiv.org/abs/0901.3426}{\texttt{arXiv:0901.3426 [hep-th]}}.

\bibitem{kastler-stora}
D.~Kastler and R.~Stora, ``{A differential geometric setting for BRS
  transformations and anomalies I}'',
\href{http://dx.doi.org/10.1016/0393-0440(86)90006-9}{\emph{J. Geom. Phys.}
  \textbf{3} (1986) 437--482}.

R.~Stora, \href{http://dx.doi.org/10.1142/9789812567147_0007}{``{From Koszul
  complexes to gauge fixing}'',} in \emph{{50 years of Yang-Mills theory}},
  G.~'t~Hooft, ed., pp.~137--167,
(World Scientific Publ., 2005).
\newblock

R.~Stora, F.~Thuillier, and J.-C. Wallet, ``\href{http://inspirehep.net/record/365824/files/RCP 25-p.139.pdf}{Algebraic structure of
  cohomological field theory models and equivariant cohomology}'', in
  \emph{{Infinite Dimensional Geometry, noncommutative Geometry, Operator
  Algebras and Fundamental Interactions}}, R.~Coquereaux, M.~Dubois-Violette,
  and P.~Flad, eds., pp.~266--297,
  {Proc. 1st Caribbean Spring School of Mathematics and Theoretical
  Physics, Saint Francois, Guadeloupe 1993},
  (World Scientific, 1995).

R.~Stora, \href{http://dx.doi.org/10.1007/978-1-4899-1801-7_11}{``{Exercises in
  equivariant cohomology}'',} in \emph{{Quantum Fields and Quantum Space
  Time}}, G.~'t~Hooft, A.~Jaffe, G.~Mack, P.~K. Mitter, and R.~Stora, eds.,
  {Proc. NATO Advanced Study Institute, Carg{\`e}se, France 1996},
  vol.~364 of \emph{{NATO Advanced Study Institutes Series, Series B}},
  pp.~265--279, (Plenum, 1997),
  \href{http://arxiv.org/abs/hep-th/9611114}{\texttt{arXiv:hep-th/9611114}}.

\bibitem{Das:2008}
A.~Das, \href{http://dx.doi.org/10.1142/6938}{\emph{{Lectures on Quantum Field
  Theory}}}, (World Scientific, 2008).

\bibitem{Frappat}
L.~Frappat, A.~Sciarrino, and P.~Sorba, \emph{{Dictionary of Lie Algebras and
  Superalgebras}}, (Academic Press, 2000).
\newblock
\href{http://arxiv.org/abs/hep-th/9607161}{\texttt{arXiv:hep-th/9607161}}.
\newblock

\bibitem{Bleecker:1981me}
D.~Bleecker, \emph{Gauge Theory And Variational Principles, Global Analysis,
  Pure and Applied, Vol. 1},
(Reading, USA: Addison-Wesley, 1981).
\newblock

\end{thebibliography}
\end{document}